\documentclass[manuscript, nonacm]{acmart}

\usepackage{tabularx}
\usepackage{xltabular}
\usepackage{array}
\usepackage{multirow}
\usepackage{adjustbox}

\newcommand{\scale}[1]{%
  \par\addvspace{\bigskipamount}%
  \noindent{\small\bfseries\MakeUppercase{#1}}%
  \par\nopagebreak\smallskip}

\newcolumntype{L}{>{\centering\arraybackslash}p{1.45cm}}
\newcommand{\lk}{{\footnotesize$\bigcirc$}}
\newenvironment{likert}[1][]
  {\def\likertnote{#1}%
   \par\noindent\textit{I encountered\ldots}\par\nopagebreak\smallskip\noindent\begin{minipage}{\linewidth}
   \footnotesize
   \begin{tabular}{@{}p{\dimexpr\linewidth-7.25cm-10\tabcolsep\relax}*{5}{L}@{}}
   \toprule
   & \textbf{Strongly disagree} & \textbf{Somewhat disagree} & \textbf{Neither agree nor disagree} & \textbf{Somewhat agree} & \textbf{Strongly agree} \\
   \midrule}
  {\bottomrule
   \end{tabular}%
   \ifx\likertnote\empty\else\par\smallskip\noindent\textit{Note:} \likertnote\fi
   \end{minipage}\par\medskip}
\newcommand{\likertitem}[1]{#1 & \lk & \lk & \lk & \lk & \lk \\}

\newenvironment{likertq}[1]
  {\par\noindent\textit{#1\ldots}\par\nopagebreak\smallskip\noindent\begin{minipage}{\linewidth}
   \footnotesize
   \begin{tabular}{@{}p{\dimexpr\linewidth-7.25cm-10\tabcolsep\relax}*{5}{L}@{}}
   \toprule
   & \textbf{Strongly disagree} & \textbf{Somewhat disagree} & \textbf{Neither agree nor disagree} & \textbf{Somewhat agree} & \textbf{Strongly agree} \\
   \midrule}
  {\bottomrule
   \end{tabular}
   \end{minipage}\par\medskip}

\newenvironment{bipolar}[1]
  {\par\noindent\textit{#1\ldots}\par\nopagebreak\smallskip\noindent\begin{minipage}{\linewidth}
   \footnotesize
   \begin{tabular}{@{}>{\raggedleft\arraybackslash}p{\dimexpr0.5\linewidth-1cm-6\tabcolsep\relax}*{5}{>{\centering\arraybackslash}p{0.4cm}}>{\raggedright\arraybackslash}p{\dimexpr0.5\linewidth-1cm-6\tabcolsep\relax}@{}}
   \toprule}
  {\bottomrule
   \end{tabular}
   \end{minipage}\par\medskip}
\newcommand{\bipolaritem}[2]{#1 & \lk & \lk & \lk & \lk & \lk & #2 \\}

\begin{document}

\title{The Serendiptionnaire: a Questionnaire for Measuring Experienced Serendipity}

\author{Brett Binst}
\email{brett.binst@vub.be}
\orcid{0000-0002-0243-9952}
\affiliation{%
  \institution{imec-SMIT, Vrije Universiteit Brussel}
  \city{Brussel}
  \country{Belgium}
}

\author{Annelien Smets}
\email{annelien.smets@vub.be}
\orcid{0000-0003-4771-7159}
\affiliation{%
  \institution{imec-SMIT, Vrije Universiteit Brussel}
  \city{Brussel}
  \country{Belgium}
}

\begin{abstract}
Serendipity is important to design for, yet it remains difficult to measure and evaluate.
No validated questionnaire exists, so studies either borrow ResQue's novelty scale, which does not measure serendipity, or build their own items, which makes findings hard to compare and synthesize.
Building on our earlier conceptualization of experienced serendipity as a fortuitous encounter with content that is both refreshing and enriching, we developed the Serendiptionnaire, a questionnaire that operationalizes our framework.
We conducted twenty-seven cognitive interviews over three iterations with respondents who varied in age, education level, country of birth, and English proficiency, while they browsed recommendations on platforms such as Instagram, TikTok, YouTube, and Spotify.
After each iteration, we revised the items until respondents interpreted them as intended.
The resulting instrument consists of modular scales that researchers can combine to target the specific flavor of serendipity they want to study or design for.
The Serendiptionnaire offers a ground truth for validating system-centered serendipity metrics and LLM-based evaluations, and it provides a nuanced evaluation of users’ experienced serendipity.
\end{abstract}

\maketitle

\section{Introduction}

Consider two user experiences with modern recommender systems (RecSys):

\bigskip

\begin{quote}
You are scrolling through your Instagram Reels; you see some Italian brainrot, followed by a video of someone imitating Haaland.
You keep scrolling, until you encounter a reel about a hidden gem in your neighborhood; a cozy new bar in a candlelit rose garden.
Perfect, you were still looking for a date idea for your latest match on Bumble.
\end{quote}

\bigskip

\begin{quote}
You are browsing your Reddit feed because you want to be up to date about the latest news.
You read some posts about the latest actions on the geopolitical stage.
Then this random video appears of someone foraging in the woods and explaining whether a handful of plants and mushrooms are toxic or not, and what happens when they are consumed.
Weirdly enough, it intrigues you, and you work your way through a few foraging communities to discover more about the topic: new interest unlocked.
\end{quote}

\bigskip

\noindent These are two very different examples of experienced serendipity, unplanned but positive encounters, when using RecSys.
Researchers agree on the importance of designing RecSys for serendipity; for instance because it stimulates taste-broadening~\citep{binst_let_2026}, flow experience~\citep{lu_sustainability_2020}, and resonance~\citep{saito_fusion-based_2022}.
However, serendipity has been characterized as a mysterious and slippery concept that is very difficult to capture~\citep{makri_making_2014}.

No agreement exists on how serendipity in RecSys should be defined~\citep{kotkov_survey_2016, ziarani_serendipity_2021}.
Most studies list unexpectedness or surprise as central~\citep{ziarani_serendipity_2021}, while others stress different elements, such as novelty~\citep{wieland_one_2021}, difficulty-to-discover~\citep{de_pessemier_comparison_2014}, or resonance~\citep{oku_user_2012}.
Its subjective nature and the conceptual ambiguity surrounding it make it hard to operationalize experienced serendipity in RecSys~\citep{kang_exploring_2025, tokutake_universal_2025}.

In \citet{binst_what_2025}, we address this conceptual ambiguity.
Drawing on interviews about specific recommendations and recommendation sessions in which users experienced serendipity, we define experienced serendipity as a fortuitous encounter with content that is both refreshing and enriching.
Rather than adding another definition, we show that each of the three core components --- fortuitous, refreshing, and enriching --- can be fulfilled through different conditions, yielding different expressions, or flavors, of experienced serendipity.
The two opening scenarios illustrate two such flavors: in the first, a fortuitous encounter resolves a latent user need; in the second, the recommendation creates a new interest.
By recognizing that experienced serendipity can express itself through different flavors, the framework bridges different notions of serendipity.

To be useful in evaluation, this conceptualization should be operationalized in a validated way.
Measuring experienced serendipity requires listening to users directly, and a carefully developed questionnaire is the best instrument for listening to users directly in a standardized way~\citep{ekstrand_behaviorism_2016}.
However, a designated and validated questionnaire for evaluating experienced serendipity is missing~\citep{kang_exploring_2025, kotkov_dark_2024, tokutake_universal_2025}.
This is problematic since a validated measurement instrument is essential to progress research~\citep{jannach_escaping_2020}.

Currently, ResQue~\citep{pu_user-centric_2011} is most often used in user centered research to ask about serendipity~\citep{binst_how_2024}.
The problem is that ResQue has no specific items asking about serendipity.
Instead, the authors suggest using the novelty scale because ``\textit{in conducting the actual user evaluation procedure, the meticulous distinction between these two words [novelty and serendipity] will cause confusion for users}''~\citep{pu_user-centric_2011}.

Although the diagnosis is true, asking users about experienced serendipity is challenging because of the complicated meaning, their solution remains problematic; by solely focusing on novelty, their operationalization cannot distinguish serendipity from a user experience that is just unfamiliar but not enriching, nor can it distinguish between the flavors illustrated by the anecdotes above.

In this paper, we develop a questionnaire that can accurately capture the nuanced and complicated meaning of experienced serendipity by conducting cognitive interviews.
Cognitive interviewing is a methodology developed to minimize response error in questionnaire design~\citep{miller_cognitive_2014, willis_cognitive_2004}.
It assumes that how participants cognitively process an item drives both their answer and their response error, so the interview works by exposing that process~\citep{willis_cognitive_2004}.
Where it diverges from what the researcher intended, the item is revised until it is understood as intended~\citep{miller_cognitive_2014, willis_cognitive_2004, willis_analysis_2015}.
The method has produced easy-to-understand questionnaires across many contexts (for instance, for measuring physical activity in older adults~\citep{heesch_what_2010}, patient reported outcomes in children~\citep{irwin_cognitive_2009}, and disability in an international population~\citep{miller_results_2011}).

We conducted and analyzed twenty-seven cognitive interviews to create what we call the \textit{Serendiptionnaire}: a validated questionnaire that allows researchers to compare findings across studies, supplies a ground truth for validating system-centered metrics, provides a rich and insightful evaluation of experienced serendipity, and can inform LLM evaluation and stimulation of experienced serendipity in RecSys.

\section{Background}

\subsection{Intended, Afforded, and Experienced Serendipity}
\label{sec:intended}

\citet{smets_intended_2025} provides a theoretical framework to overcome the apparent paradox of designing technologies for the unplanned by distinguishing between three levels of serendipity:
\begin{itemize}
  \item \textit{Intended serendipity}: This level concerns the strategic goal a design should serve.
  By setting the goal, it determines which flavors of serendipity are relevant.
  \item \textit{Afforded serendipity}: This level concerns the capacities of the system to afford experienced serendipity.
  \item \textit{Experienced serendipity}: This level concerns serendipity as it is actually experienced and evaluated, and represents the perspective of the end user.
\end{itemize}

\begin{figure}[htbp]
  \centering
  \includegraphics[width=0.6\linewidth]{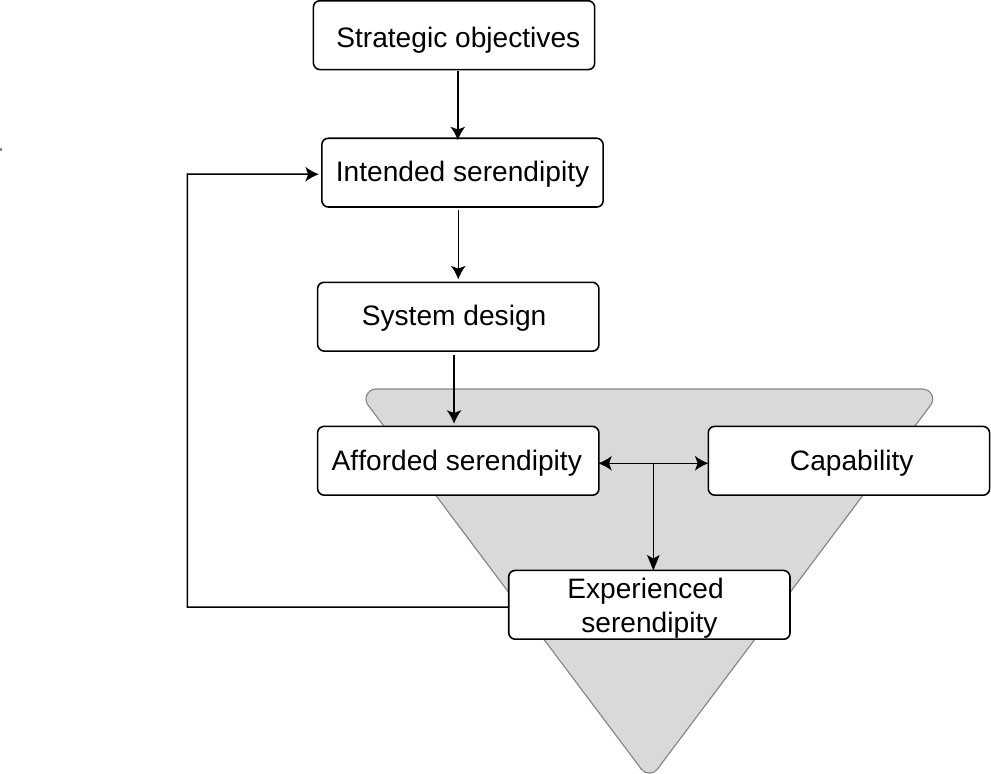}
  \caption{Theoretical model of serendipity distinguishing three intertwined levels: intended, afforded, and experienced serendipity (adapted from \citet{smets_intended_2025}).}
  \label{fig:framework}
\end{figure}

The framework acknowledges that experienced serendipity comes in different flavors~\citep{smets_intended_2025}, and the levels inherit this plurality from one another.
Different flavors correspond to different reasons to design for serendipity.
For example, a design aimed at diversifying consumption (\textit{intended serendipity}) targets taste-broadening serendipity, whereas a design supporting a user with a specific goal (for instance, finding tasty recipes that follow a specific diet) would target relevant-but-difficult-to-find serendipity.
The best practices at the afforded level likewise depend on the intended level, since the most suitable feature or algorithm depends on the specific design goal.
The three levels are therefore intertwined as shown in Figure~\ref{fig:framework}.

The framework overcomes the paradox of designing for the unplanned because it is not deterministic; you can design for afforded serendipity, but this will only result in experienced serendipity when the capability to actualize the affordance is present in the user~\citep{gaver_technology_1991}.
For example, diversifying movie recommendations will only result in taste-broadening serendipity if the user is actually open to discover something new, and intrigued by the unusual recommendations~\citep{liang_recommender_2019}.

\subsection{Experienced Serendipity}
\label{sec:experienced}

\begin{figure}[H]
  \centering
  \includegraphics[width=0.8\linewidth]{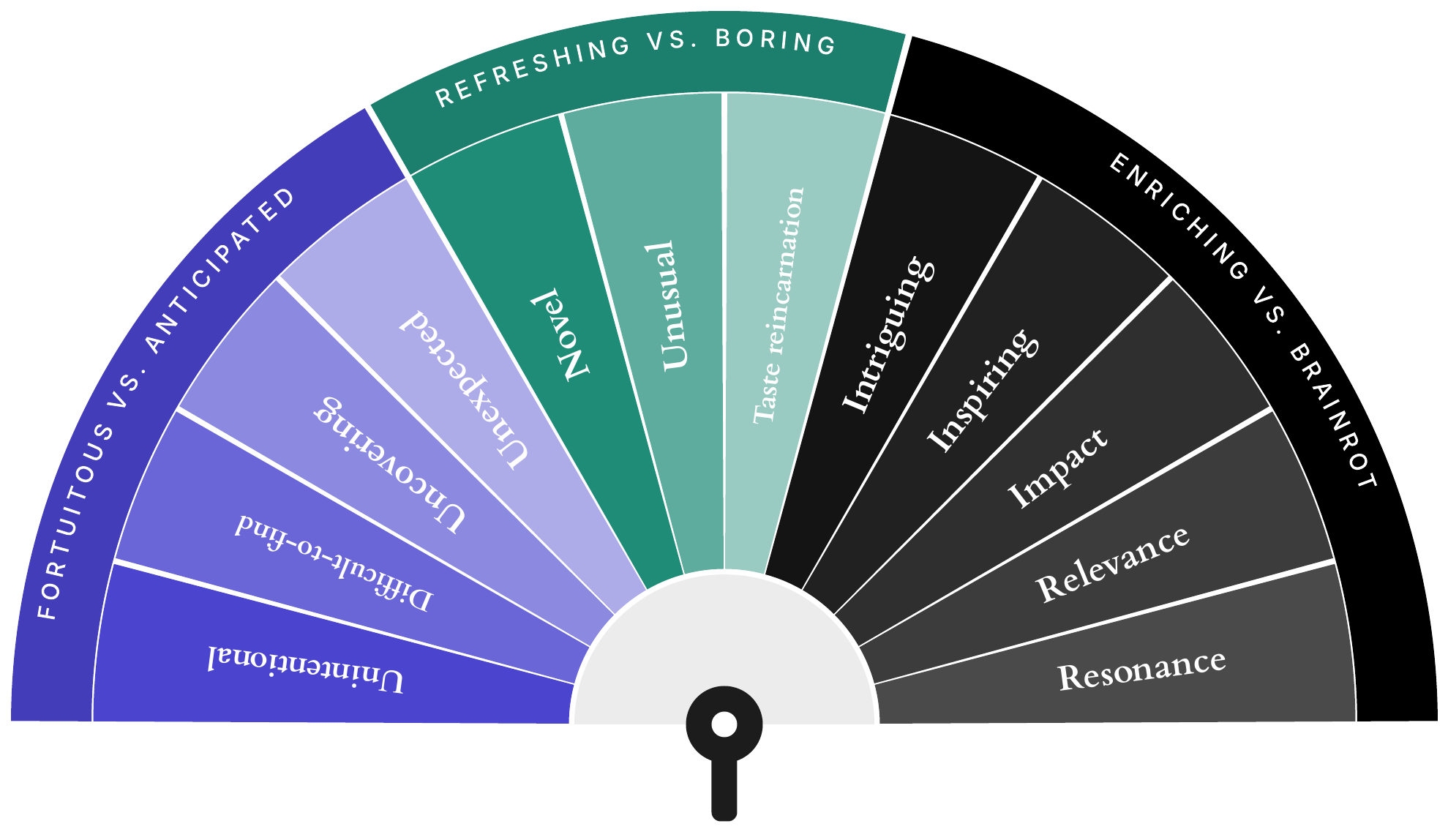}
  \caption{The three core components of experienced serendipity and the conditions through which each can be fulfilled.}
  \label{fig:fan}
\end{figure}

The Serendiptionnaire measures experienced serendipity.
In \citet{binst_what_2025}, we conducted a grounded theory analysis of when users experience serendipity while interacting with RecSys.
We found that experienced serendipity has three core components: a fortuitous encounter with content that is both refreshing and enriching.
Each component can be realized in more than one way, as Figure~\ref{fig:fan} shows.
Experienced serendipity therefore expresses itself in different ways, referred to as different flavors of serendipity.

Flavors are distinguished by which combinations of conditions they require.
Some are partly defined by conditions that are not fulfilled, for example taste-deepening involves content that is usual for the user, yet remains refreshing because it is uncovering.
Underneath, we shortly describe what is meant with each component and condition.

\subsubsection{Fortuitous}
Concerns how content is encountered.
Its conditions describe different ways in which an encounter can be fortuitous.

\begin{description}
  \item[Unintentional:] the recommendation is encountered rather than sought.
  It implies a vague user intention, or an encounter that diverges from the intention the user held.
  \item[Unexpected:] the user perceives the encountered content as unexpected because they did not expect to encounter this item, or because they are surprised by the content of the item.
  \item[Difficult-to-find:] the user perceives the encounter as facilitated by the system: without this platform, the item would have been harder to come across.
  It also covers the perception of an item as a hidden gem, something interesting that the user did not know yet or had overlooked until now.
  \item[Uncovering:] the encounter reveals something the user did not previously see.
  This may mean approaching an item from a new perspective, or learning something new about oneself.
\end{description}

\subsubsection{Refreshing}
Concerns the relation between the encountered content and the user's past.

\begin{description}
  \item[Novel:] the item is new to the user.
  Novelty can express itself to different degrees: an item the user did not know about yet, an item the user did know about but did not yet consume (watch, listen, read), or an item that was already consumed by the user.
  \item[Unusual:] the item differs from what a user normally consumes, such as a genre they never listen to.
  It is important to distinguish novelty from unusualness; a song can be novel but perfectly aligned with the listener's typical preferences.
  Conversely, a song can be familiar but strongly different from the listener's typical preferences.
  \item[Taste-reincarnation:] the content reconnects the user with their past and/or stimulates nostalgia, whether by rediscovering material from childhood or by encountering something new that recalls an earlier period.
\end{description}

\newpage

\subsubsection{Enriching}
Concerns the positive effect of the encounter.

\begin{description}
  \item[Intriguing:] the content sparks curiosity or interest, either by opening up a new interest by introducing the user to an unusual item or by deepening an existing interest.
  \item[Energizing:] the user encounters something that activates them, whether motivating them to do a specific action or making them feel recharged because the content disconnects them from frustrations.
  \item[Inspiring:] the user's imagination or ideas are enriched.
  It implies that the user has the impression of having more options of acting than before the encounter.
  This can mean increased creativity by introducing the user to a new technique for painting for example or stimulating the user's imagination and in that way enriching their perception about a certain topic.
  \item[Impact:] the encounter might have a positive influence on the user's life.
  For example, it could mean that the item helps the user in solving a problem they were facing (i.e., useful), but it could also mean that it drastically changes the user's perception about something and therefore changes their life in an enriching way.
  \item[Empowering:] the encounter raises the user's self-esteem through affirmation or an increased sense of agency.
  \item[Usefulness:] the encounter is instrumental to a goal the user holds, whether a current problem or an anticipated one.
  \item[Resonance:] the encounter produces either a positive emotional response or a sense that the item is meaningful and aligned with the user's values.
\end{description}

\subsection{Measuring Serendipity}

Not all questionnaires that operationalize serendipity operate on the same level.
For instance, \citet{mccay-peet_examination_2015} developed a questionnaire to measure what we call \textbf{afforded serendipity}.
The questionnaire operationalizes \citeauthor{bjorneborn_design_2010}'s~\citep{bjorneborn_design_2010} serendipity affordances which he discovered when observing people serendipitously discovering books in libraries.
\citet{kim_development_2025} on the other hand develop a questionnaire measuring the individual propensity (which aligns with \textbf{capability} in Figure~\ref{fig:framework}) to experience serendipity.
Also this questionnaire has a strong theoretical foundation; it is grounded in the process model of information encountering~\citep{erdelez_investigation_2004} and serendipity~\citep{erdelez_information_2020}.

In RecSys research, ResQue~\citep{pu_user-centric_2011} is often used for measuring \textbf{experienced serendipity} (for example, \citep{afridi_transparency_2019, cai_critiquing_2021, chen_how_2019, taramigkou_supporting_2017}).
However, ResQue is a general framework for measuring user experience in RecSys, and not developed for measuring serendipity.
Actually, ResQue does not even include a designated item for measuring serendipity.
Instead, \citet{pu_user-centric_2011} suggest using the novelty scale as a proxy for measuring serendipity since ``\textit{in conducting the actual user evaluation procedure, the meticulous distinction between these two words [novelty and serendipity] will cause confusion for users}''~\citep{pu_user-centric_2011}.
However, novelty is one condition among several, and cannot separate a serendipitous encounter from an encounter that is merely unfamiliar, nor distinguish one flavor of serendipity from another.
That a validated scale for a different construct is routinely pressed into this role indicates the absence of a validated instrument for measuring experienced serendipity itself.

\newpage

Besides ResQue, several studies operationalized their own notion of experienced serendipity through questionnaire items.
Table~\ref{tab:operationalizations} shows an overview.

{\small
\begin{xltabular}{\linewidth}{@{}p{3.2cm}Xp{3.6cm}@{}}
  \caption{Questionnaire items used in previous studies to operationalize experienced serendipity.}
  \label{tab:operationalizations} \\
  \toprule
  \textbf{Authors} & \textbf{Questions Asked} & \textbf{Dimensions Captured} \\
  \midrule
  \endfirsthead
  \multicolumn{3}{@{}l}{\textit{Table~\ref{tab:operationalizations} continued}} \\
  \toprule
  \textbf{Authors} & \textbf{Questions Asked} & \textbf{Dimensions Captured} \\
  \midrule
  \endhead
  \bottomrule
  \endlastfoot
  \citet{pu_user-centric_2011} & The items recommended to me are novel; The recommender system helped me discover new products & Novelty \\
  \addlinespace
  \citet{oku_user_2012} & I did not know this book; I have been interested in this book before the system presented it to me; This book excited my interest for the first time; I think that I could not find this book by myself & Novelty; prior interest; intriguing; difficult-to-find \\
  \addlinespace
  \citet{holldobler_serendipity_2015} & This article is relevant regarding the source article; I am positively surprised by this article, I am glad I found it & Relevance; surprise; positive affect \\
  \addlinespace
  \citet{guy_islands_2015} & How interesting is the news item for you?; How surprising is the news item for you? & Interest; Surprise \\
  \addlinespace
  \citet{maccatrozzo_sirup_2017} & I did not think of this TV programme, but it seems interesting to me; This TV programme does not seem interesting to me; I am surprised to get this TV programme recommended; This recommendation fits my personal preference & Novelty; Interest; Surprise; Relevance \\
  \addlinespace
  \citet{kotkov_investigating_2018} & The first time I heard of this movie was when MovieLens suggested it to me; MovieLens influenced my decision to watch this movie; I expected to enjoy this movie before watching it for the first time; This is the type of movie I would not normally discover on my own; I need a recommender system like MovieLens to find movies like this one; This movie is different (e.g., in style, genre, topic) from the movies I usually watch; I was (or, would have been) surprised that MovieLens picked this movie to recommend to me & Novelty; Influence; Difficult-to-find; Unusual; Surprise; Satisfaction; Preference broadening \\
  \addlinespace
  \citet{ningxia_impacts_2020} & The item recommended to me is a pleasant surprise & Surprise; positive affect \\
\end{xltabular}
}

These operationalizations have many differences.
Clearly, they measure different user experiences, yet they state the same conclusion: the system they developed stimulated serendipity.
However, comparison and synthesis across studies becomes difficult because of the different meanings attached to serendipity, which is why the conceptual and operational ambiguity surrounding serendipity is regarded as one of the primary bottlenecks to progress in research on serendipity~\citep{kotkov_challenges_2016, ziarani_serendipity_2021}.

Read together however these instruments do point to a common ground; they converge on the structure of experienced serendipity as a polythetic, composite concept, that is, consisting of a variety of criteria, and that can express itself in different ways.
For example, \citet{oku_user_2012} distinguish between users serendipitously discovering a new book and serendipitously getting excited in a known but unread book for the first time.
\citet{kotkov_investigating_2018} go even further by distinguishing between eight different expressions of serendipity based on the questions listed in Table~\ref{tab:operationalizations}.
The fragmentation in operationalizations is a symptom of this polythetic composite structure; different operationalizations are measuring different expressions of experienced serendipity.
Our conceptualization of experienced serendipity described in Section~\ref{sec:experienced} accommodates this structure.
Now, it is time to operationalize this conceptualization into a questionnaire that is able to measure these different notions of experienced serendipity.

\section{Methodology}

\subsection{Cognitive Interviewing Approach}

A validated questionnaire measuring experienced serendipity requires items that respondents read the way we meant them.
Cognitive interviewing is a method to maximize the likelihood of the items being interpreted as intended~\citep{willis_cognitive_2004, willis_analysis_2015}.
Below, we describe how we conducted the cognitive interviews to create the Serendiptionnaire.

\subsection{Sampling Approach}
\label{sec:sampling}

To ensure the Serendiptionnaire captures the experienced serendipity of a diversity of users, we purposefully sampled across three interview iterations between May 2025 and May 2026~\citep{campbell_purposive_2020, strauss_basics_2015}.
Four characteristics guided sampling: education level, age, country of birth, and English proficiency.
Additionally, we developed the Serendiptionnaire across a range of platforms.

As recommended by \citet{willis_analysis_2015}, we conducted the cognitive interviews in three iterations of analyzing feedback, refining questionnaire items, and testing the revised instrument with a new sample, and included nine respondents per iteration.
The sample is described in Table~\ref{tab:participants}.

\begin{table}[htbp]
  \caption{Participants and the platforms used during the cognitive interviews.}
  \label{tab:participants}
  \footnotesize
  \setlength{\tabcolsep}{4pt}
  \begin{tabular*}{\linewidth}{@{\extracolsep{\fill}}lllllll@{}}
    \toprule
    \textbf{Participant} & \textbf{Platform} & \textbf{English proficiency*} & \textbf{Highest attained degree} & \textbf{Country of birth} & \textbf{Age} & \textbf{Gender} \\
    \midrule
    \multicolumn{7}{@{}l}{\textit{Iteration 1}} \\
    P1 & Instagram (reels) & / & Technical secondary education & Belgium & 26 & M \\
    P2 & TikTok & B1/B2: 75\% & Technical secondary education & Belgium & 18 & M \\
    P3 & TikTok & C1/C2: 95\% & Master's & Spain & 28 & W \\
    P4 & LinkedIn & B1/B2: 65\% & Vocational secondary education & Belgium & 60 & M \\
    P5 & RedNote & B1/B2: 85\% & Master's & China & 37 & W \\
    P6 & Instagram (explore) & B1/B2: 80\% & Master's & Belgium & 44 & W \\
    P7 & Zalando & C1/C2: 90\% & Master's & Indonesia & 28 & M \\
    P8 & Zalando & C1/C2: 95\% & Master's & Poland & 28 & W \\
    P9 & Bol.com & C1/C2: 95\% & Bachelor & Belgium & 33 & W \\
    \midrule
    \multicolumn{7}{@{}l}{\textit{Iteration 2}} \\
    P10 & Instagram reels & B1/B2: 85\% & Bachelor & Italy & 24 & W \\
    P11 & YouTube & B1/B2: 85\% & PhD & Costa Rica & 35 & M \\
    P12 & Bilibili & B1/B2: 65\% & Master's & China & 39 & M \\
    P13 & Spotify & C1/C2: 100\% & PhD & USA & 31 & W \\
    P14 & Instagram reels & B1/B2: 80\% & Master & Belgium & 27 & M \\
    P15 & Instagram reels & C1/C2: 95\% & Bachelor & Colombia & 27 & W \\
    P16 & Spotify & B1/B2: 75\% & PhD & Kenya & 37 & M \\
    P17 & YouTube & C1/C2: 95\% & Master & India & 38 & M \\
    P18 & Instagram Explore & C1/C2: 90\% & Bachelor & Ghana & 29 & M \\
    \midrule
    \multicolumn{7}{@{}l}{\textit{Iteration 3}} \\
    P19 & YouTube shorts & C1/C2: 95\% & Office Management & Germany & 32 & W \\
    P20 & Spotify & C1/C2: 100\% & PhD & Sweden & 69 & M \\
    P21 & Instagram Reels & B1/B2: 75\% & High School Degree & Philippines & 21 & W \\
    P22 & YouTube & C1/C2: 90\% & Master & India & 24 & M \\
    P23 & LinkedIn & C1/C2: 90\% & Master & Libya & 33 & M \\
    P24 & Threads & C1/C2: 95\% & Doctor in Law & USA & 40 & M \\
    P25 & Facebook & C1/C2: 95\% & Technical Secondary Education & Belgium & 57 & W \\
    P26 & Instagram Reels & C1/C2: 95\% & Master & Italy & 26 & W \\
    P27 & TikTok & B1/B2: 70\% & Bachelor & Belgium & 27 & W \\
    \bottomrule
  \end{tabular*}

  \smallskip
  \raggedright\footnotesize\textit{Note:} *English proficiency was measured using EF SET's English practice test.
\end{table}

\subsection{Interview Procedure}

Cognitive interviewing reveals how respondents understand and answer questionnaire items, which makes it central to questionnaire validation~\citep{willis_cognitive_2004, willis_analysis_2015}.
\citet{willis_analysis_2015} distinguishes a descriptive from a reparative approach.
The descriptive approach documents how respondents interpret and answer items.
The reparative approach goes further and moves that interpretation toward the researcher's intended meaning.
In this work, we adopt the reparative approach, since the aim was to align respondents' interpretations with the conceptual framework of experienced serendipity~\citep{binst_what_2025}.

Interviews took place in person where possible and by video call with respondents living abroad.
Each interview followed four steps.
First, the interviewer explained the procedure and the informed consent.
Second, respondents took the English proficiency test.
Third, respondents were asked to have a normal session on a familiar platform leveraging a recommender system (for example, scrolling through instagram reels) while the interviewer observed; watching the session made the reasoning behind their answers easier to follow.
Table~\ref{tab:participants} lists the systems used.
Fourth, the cognitive interview was done while participants filled in the Serendiptionnaire.

Cognitive interviewing consists of two main techniques: think-aloud and probing~\citep{willis_cognitive_2004}.
Think-aloud requires respondents to say aloud what they read, how they interpret the question, and the cognitive process that results in their answer.
During verbal probing the interviewer asks specific probes to get more targeted insight into their reasoning.
Both are complementary; if the respondent's automatic reflections are not satisfactory for assessing how the question is understood and answered, verbal probing is used to get more insight.

We asked respondents to think aloud while completing the questionnaire; explaining how they interpret the item, what they do not understand, and what they are thinking about when answering the item.
When necessary, we followed up with additional probing to gather richer feedback.
Elaborative probes encouraged respondents to explain their reasoning, for instance: ``Why did you choose this answer?'' or when apparently conflicting answers were given, ``Why did you answer this item differently compared to this one?''
Meaning-oriented probes explored respondents' interpretations of key terms, such as ``What does `intriguing' mean to you?''
Furthermore, we want the Serendiptionnaire to be inclusive and sensitive for ``micro-serendipities''~\citep{bogers_micro-serendipity_2013}, i.e., small everyday occurrences of serendipity.
This requires hypothetical probes, for example: ``In what situation would you select `Strongly agree' to this question?'' in order to increase the sensitivity of the Serendiptionnaire.
Finally, it is also important that the questionnaire is understandable for a diversity of people and to avoid what \citet{fricker_epistemic_2007} called hermeneutical injustice, a situation in which people experience unfair obstacles in communication~\citep{medina_varieties_2017}.
Overcoming hermeneutical injustice requires attention to social inequality in understanding questionnaire items~\citep{binst_bourdieusian_2025}.
Therefore, we purposely sampled for different levels of English proficiency and education level and used evaluative probes to assess the clarity and difficulty of the items, such as ``\textit{Was this question difficult to answer?}'' when respondents paused for a longer time.

\subsection{Analysis}

We transcribed all interviews using Scribewave.
Subsequently, transcripts were coded in MAXQDA.
The analysis was focused on different elements of the questionnaire: the general introduction, the introduction to the different question matrices, the question format, and the individual items.

We coded our data both deductive and inductive.
For deductive coding, we draw on the Question Appraisal System~\citep{willis_question_1999}; a framework to appraise the strengths and weaknesses of questionnaire items (see Table~\ref{tab:qas}).
We used inductive coding to signal problems with and strengths of the items that are not included in the Question Appraisal System.
Furthermore, inductive coding was essential for understanding respondents' specific interpretation of the questionnaire items.

{\small
\begin{xltabular}{\linewidth}{@{}>{\raggedright\arraybackslash}p{3.9cm}X@{}}
  \caption{Codes of the Question Appraisal System~\citep{willis_question_1999} used for deductive coding.}
  \label{tab:qas} \\
  \toprule
  \textbf{Code} & \textbf{Used when\ldots} \\
  \midrule
  \endfirsthead
  \multicolumn{2}{@{}l}{\textit{Table~\ref{tab:qas} continued}} \\
  \toprule
  \textbf{Code} & \textbf{Used when\ldots} \\
  \midrule
  \endhead
  \bottomrule
  \multicolumn{2}{@{}p{\linewidth}@{}}{\footnotesize\textit{Note:} Codes marked $^\dagger$ were added by us. Not all codes of \citet{willis_question_1999} are included because not all are relevant for the Serendiptionnaire (e.g., codes on Knowledge/Memory).} \\
  \endlastfoot
  \multicolumn{2}{@{}p{\linewidth}@{}}{\textbf{Instructions:} \textit{problems with any introductions, instructions, or explanations, from the respondent's point of view}} \\
  \addlinespace[2pt]
  Conflicting or inaccurate & introductions, instructions, or explanations are conflicting or inaccurate \\
  Complicated & introductions, instructions, or explanations are complicated \\
  \midrule
  \multicolumn{2}{@{}p{\linewidth}@{}}{\textbf{Clarity:} \textit{problems in communicating the intent or meaning of the question to the respondent}} \\
  \addlinespace[2pt]
  Wording & the question is lengthy, awkward, ungrammatical, or contains complicated syntax \\
  Technical terms & technical terms are undefined, unclear, or complex \\
  Vague & there are multiple ways to interpret the question or to decide what is to be included or excluded \\
  Reference periods & reference periods are missing, not well specified, or in conflict \\
  \midrule
  \multicolumn{2}{@{}p{\linewidth}@{}}{\textbf{Assumptions:} \textit{problems with the assumptions made or the underlying logic}} \\
  \addlinespace[2pt]
  Inappropriate assumptions & inappropriate assumptions are made about the respondent or about their living situation \\
  Assumes constant behavior & the question assumes constant behavior or experiences for situations that vary \\
  Double-barreled & the question contains more than one implicit question \\
  \midrule
  \multicolumn{2}{@{}p{\linewidth}@{}}{\textbf{Sensitivity/bias:} \textit{sensitive nature or wording of the question, and bias}} \\
  \addlinespace[2pt]
  Sensitive content (general) & the question asks about a topic that is embarrassing, very private, or that involves illegal behavior \\
  Sensitive wording (specific) & the general topic is sensitive and the wording should be improved to minimize sensitivity \\
  Socially acceptable & a socially acceptable response is implied by the question \\
  \midrule
  \multicolumn{2}{@{}p{\linewidth}@{}}{\textbf{Response categories:} \textit{adequacy of the range of responses to be recorded}} \\
  \addlinespace[2pt]
  Open-ended question & an open-ended question is inappropriate or difficult \\
  Mismatch & the question and the response categories do not match \\
  Technical terms & technical terms in the response categories are undefined, unclear, or complex \\
  Vague & response categories are subject to multiple interpretations \\
  Overlapping & response categories overlap \\
  Missing & eligible responses are missing from the response categories \\
  Illogical order & response categories are in an illogical order \\
  \midrule
  \multicolumn{2}{@{}p{\linewidth}@{}}{\textbf{Other problems:} \textit{problems not identified above}} \\
  \addlinespace[2pt]
  Other & there are other problems not previously identified \\
  Not relevant$^\dagger$ & the item is perceived as irrelevant \\
  Difficult to answer$^\dagger$ & the item is difficult to answer \\
\end{xltabular}
}

As explained in Section~\ref{sec:sampling}, the Serendiptionnaire has been developed iteratively.
Therefore, the analysis was also iterative.
Every iteration entailed 1) coding the transcripts, 2) listing relevant information and quotes about the items per scale for every participant, 3) summarizing this information per scale, and finally 4) adapting the scale based on the insights gained.
This process resulted in the final version of the Serendiptionnaire which we will discuss in the following section.

\section{The Serendiptionnaire}
\label{sec:serendiptionnaire}

This section discusses the final version of the Serendiptionnaire and how it evolved throughout the three iterations.
The final version of the Serendiptionnaire is included in full in Appendix~\ref{app:serendiptionnaire}, and the three previous versions which were tested throughout the iterations can be found in Appendices~\ref{app:iteration1}--\ref{app:iteration3}.

\subsection{Introduction to the Serendiptionnaire}

At the start of the Serendiptionnaire, participants read the following introduction:

\begin{quote}
\textit{In this questionnaire, we are interested in how you feel about the content you encountered \textbf{during the session you just finished}.
There are no right or wrong answers, so please don't overthink the questions, just answer based on what comes to mind first.}
\end{quote}

The first iteration had no introduction.
However, we noticed that an introduction could help respondents understand the intended meaning.
So in the second iteration, we added an introduction, which we shortened in the third iteration.
The introduction stresses that we are interested in the session the respondent just had, and in their feeling about the content they encountered during that session.
Respondents who read the introduction understood it as intended, for instance P20: ``\textit{I actually noticed that you are using feel. So actually you're already kind of framing me to think about my emotions.}''
However, many respondents do not pay attention to the introduction; even when the researcher is sitting next to them.

\subsection{Fortuitous vs.\ Anticipated}

\scale{Unintended}

\begin{likert}
\likertitem{\textbf{something} I did not specifically search for during this session}
\likertitem{\textbf{something} I was not specifically looking for during the session}
\end{likert}

\textbf{Development:} Only the most specific items were kept in the final scale.
The ``did not search for'' item was made more specific because it covers a distinct element of the condition: ``looking for'' can include browsing, whereas ``search for'' lowers the threshold to anything not actively sought.

\textbf{Interpretation:} The most frequent, and hardest to solve, misunderstanding in this scale is that the specific encounter (for example, a funny video) is confused with the goal of the session (for example, browsing and entertainment).
From this goal, the user often evaluates the session as intended (for example, the goal was to scroll for entertainment; I saw a funny video. So, the session was as intended).
When thinking about the specific items, conversely, the user evaluates the encounters as unintended (For example, the user did not plan to find exactly that funny video).
Therefore, it will be especially important to specify the referent (is the session the level of analysis or individual items) when measuring this condition.
The more specific items indeed performed better, as P20 noted: ``\textit{the fourth one [something I was not specifically looking for during the session] is also a good one, because there is even more specification about during the session.}''

\scale{Difficult to Find}

\begin{likert}
\likertitem{a hidden gem (\textbf{something} interesting but that I did not know yet, or overlooked until now)}
\likertitem{\textbf{something} that might have been more difficult to discover without the help of this platform}
\end{likert}

\begin{likert}
\likertitem{\textbf{mainly} content that I would have easily discovered without the platform}
\end{likert}

\textbf{Development:} Respondents with weaker English did not understand hidden gem in earlier iterations.
Hence, we added an explanation and found that this did clarify the intended meaning.

\textbf{Interpretation:} The items were understood as intended by all respondents in the last iteration except P27 whose English proficiency was among the lowest in the sample.
However, the ``hidden gem'' item was easy to understand for her.
Furthermore, we observe a difference between how ``hidden gem'' and ``easy to discover without the platform'' are interpreted.
A respondent may encounter a hidden gem that would nonetheless have been easy to find elsewhere.
For example, P14 stated that he encountered a hidden gem when scrolling through Instagram Reels: ``\textit{If we're talking about places to eat or drink something, yes, somewhat agree. That is a hidden gem.}''
However, he stated it would have been easy to find elsewhere: ``\textit{Would I find these topics easily on different platforms? On YouTube I would, so somewhat agree.}''
The converse also occurred: content perceived as difficult to find without the platform is not necessarily perceived as a hidden gem.

\scale{Unexpected}

\begin{likert}
\likertitem{\textbf{something} that surprised me}
\likertitem{\textbf{something} unexpected}
\end{likert}

\begin{likert}
\likertitem{\textbf{nothing} surprising to me}
\likertitem{\textbf{only} content that I expected to find}
\end{likert}

\textbf{Development:} In previous iterations we asked about surprise, unexpectedness, and predictability.
Predictability was dropped because it was understood as a synonym of expected.

\textbf{Interpretation:} both surprise and unexpected are interpreted as intended.
Often respondents interpreted them as synonyms: ``\textit{Something surprising. If it's unexpected, then I assume it's also surprising, so\ldots}''
However, sometimes their interpretation differed as P25 illustrates: ``\textit{Yes, for example I follow this hotel in Außen, and they always share these surprising pictures, but for me that is no surprise, but it is unexpected sometimes, those pictures.}''

``Expected to find'' is ambiguous according to P20: ``\textit{only content I expected to find. I would have difficulties with that as well to answer. Because what does it mean to expect to find? I expect to find, if I use time enough on Spotify, I expect to find stuff I did not know about beforehand}'' implying that this item should also be adapted to the specific research context.

Respondents also differed in what they compared the encounter against.
Most judged it against their previous interactions on the platform, as P23 did: ``\textit{there were a couple of posts there that I was like, `What is this doing here?' [\ldots] that's a surprise because that's not my usual content.}''
Others judged the content itself, as P21 did: ``\textit{a Reel of a dog standing in front of a car while everybody else is already crossing the pedestrian lane [\ldots] that surprised me in a way 'cause it's not the usual behavior of dogs.}''

\scale{Uncovering}

\begin{likert}[this scale cross-cuts the fortuitous and refreshing component.]
\likertitem{\textbf{something} that helped me realize something new}
\likertitem{\textbf{something} that I did not know could be useful to me}
\likertitem{\textbf{something} that made me aware about something I did not know about before}
\likertitem{\textbf{something} that made me notice details I did not notice before}
\end{likert}

\textbf{Development:} Earlier iterations asked about personal revelations (``content that helped me realize something new about my life,'' ``something I didn't know I was interested in'').
These were seldom applicable.
Uncovering is broader than just uncovering something about the person, it could also be about uncovering something about the recommended items, or the item could also uncover other things that are less significant than personal revelations.
Therefore, the items were rephrased to be sensitive to these other expressions.

\textbf{Interpretation:} Respondents perceived both hierarchy and distinction among the final items.
For instance, ``I did not know could be useful to me'' differs from the other items in the sense that there is an extra criteria; the encounter needs to have a perceived use.
There is also a distinction between ``made me notice details I did not notice before'' and the other items in the sense that this item specifically focuses on details, and is hence more suited for measuring the dimension of uncovering in familiar items.
Finally, ``made me aware'' is read as a small change in perception while ``realize'' is perceived as more fundamental.
Two items were hard to understand for some respondents: ``something that made me aware about something I did not know about before'' for several, and ``something that made me notice details I did not notice before'' for P25, plausibly because of the abstractness of \textit{something}.
Specification should resolve both.
However, in general the items are interpreted as intended, for example P22 reflects on his session on YouTube: ``\textit{Something that made me aware about something. Yeah, I would say strongly agree because I did definitely not know about graffitis or the fact that people are painting video game worlds and this kind of thing. So I think that made me aware of that. So that's a good new area.}''

\subsection{Refreshing vs.\ Boring}

\scale{Novelty}

\begin{likert}
\likertitem{\textbf{at least one item} I had not seen before}
\likertitem{\textbf{at least one item} I did not know about yet}
\end{likert}

\begin{likert}
\likertitem{\textbf{mainly} items I had seen before}
\likertitem{\textbf{mainly} items I knew about already}
\end{likert}

\textbf{Development:} Earlier iterations used novel, familiar/unfamiliar, and new.
These were dropped: novel was not understood by all respondents, and familiar and new were too often read as synonyms of unusual.
``Not seen before'' and ``did not know about yet'' capture the intended meaning while separating this scale from unusual.

\textbf{Interpretation:} Respondents perceived the items of this scale as concrete, specific, and easy to answer, as P20 states: ``\textit{Okay. I encountered only items\ldots\ Here it's easier for me. I can relate more to some kind of practicality about it or encountered at least.}''
P24, a lawyer, illustrates the intended reading: ``\textit{I encountered only items that I'd seen before. Well, I would disagree with that. You know, like those court cases, I didn't see those actual documents before.}''

However, also here it will be important to replace ``items'' with a more specific level of analysis to make the scale as clear as possible.
As P20 states: ``\textit{what is the degree of did not know? I don't know about this series. I don't know about Albanian history? So, here I would be a little uncertain actually.}''

\scale{Unusual}

\begin{likert}
\likertitem{\textbf{something} different from the type of content I normally consume}
\likertitem{\textbf{something} different from my typical preferences}
\end{likert}

\begin{likert}
\likertitem{\textbf{mainly} the same genre of content I normally consume}
\end{likert}

\newpage

\textbf{Development:} Items about content that ``broadened my horizon'' or ``expanded my perspective'' were dropped because they imply that the content is also enriching, which the instrument measures separately.
P25 read ``outside my typical preferences'' as the algorithm showing her content she would prefer not to see, so the item was changed to ``different from my typical preferences.''
A second item containing preferences was dropped since it was too similar to the first.

\textbf{Interpretation:} Apart from P25's misreading, which no other participant shared, the items were interpreted as intended, as in P22's response: ``\textit{I encountered only the same genre of content. Yeah, absolutely \ldots\ it's the same genre, different topics. I encountered something different from the type of content I normally consume. Disagree or strongly disagree.}''

\scale{Taste-Reincarnation}

\begin{likert}
\likertitem{\textbf{something} that feels nostalgic to me}
\likertitem{\textbf{something} that reconnected me to something that I previously liked}
\end{likert}

\begin{likert}
\likertitem{\textbf{something} from my personal past that I prefer to avoid}
\likertitem{\textbf{something} that reduced my interest in something I previously liked}
\end{likert}

\textbf{Development:} Wordings such as ``revived previous passions'' were not understood and were dropped after the first iteration.
Items about rediscovering one's past self were deleted, since most respondents did not find them applicable.

\textbf{Interpretation:} The persistent problem in the later iterations was the ambiguity of \textit{the past}.
For example, P22 read \textit{nostalgic} and \textit{past} as referring simply to content he had already seen.
P26 reflected on this ambiguity explicitly: ``\textit{what do you define as past? 'Cause if it's content that I've shared before and that's somewhat a stream, then yeah,.}''
The same vagueness appeared in the previous iteration, as P12 illustrates: ``\textit{past can be my past experience, my past life. It also can be history, right?}''

Two changes followed.
The nostalgia item was rephrased to ``something that feels nostalgic to me,'' which rephrases the judgment in the respondent's subjective perception rather than in an objective past.
Furthermore, the reconnection item was rephrased to ``reconnected me to something that I previously liked,'' similarly to the negative item of the scale, which was already read as intended like P24 illustrates: ``\textit{I'm really interested in this topic, and then I look at this content, and I'm like, `Eh, I don't really wanna read about that anymore.'}''

\subsection{Enriching vs.\ Brainrot}

\scale{Intrigue}

\begin{likert}
\likertitem{\textbf{something} that sparked my interest}
\likertitem{\textbf{something} that stimulated my curiosity}
\end{likert}

\begin{likert}
\likertitem{\textbf{mainly} content that failed to grab my attention}
\end{likert}

\textbf{Development:} ``Something that intrigued me'' and ``left me indifferent'' were dropped because \textit{intrigued} and \textit{indifferent} were not understood by some respondents.

\textbf{Interpretation:} The remaining items were understood as intended as P23 illustrates: ``\textit{Something that sparked my interest. Yeah. There were a few posts where I'm like, `Hmm, interesting.' Something that stimulated my curiosity. Yes. Agree. And yes, like this post is a good example that I'm gonna read after.}''

Some respondents differ in their interpretation of ``stimulated my curiosity'' and ``sparked my interest.''
P26 and P20 state that interest is already acquired taste while curiosity is taste-broadening, while P24 states the opposite.
For other respondents, either curiosity or interest has a higher threshold and hence is more difficult to achieve.
However, this is not problematic since both interpretations fit the interpreted meaning.

\scale{Energy}

\begin{likert}
\likertitem{\textbf{something} that recharged me}
\likertitem{\textbf{something} that gave me energy}
\likertitem{\textbf{something} that motivates me to \ldots}
\end{likert}

\begin{likert}
\likertitem{\textbf{something} that lowered my energy}
\likertitem{\textbf{something} that left me feeling like a couch potato (lazy and without energy).}
\end{likert}

\textbf{Development:} The item about motivation is reintroduced in the final iteration as a response to the interpretation of energizing as similar to motivation, but more general.
However, it should be specified since motivation in general misses a direction.
Therefore, we changed the item into a template ``Motivated me to \ldots?'' which should be specified (for example, to cook healthy, or to save energy).
Furthermore, a definition was added to ``couch potato'' since several respondents were not sure what it meant.
For instance, P18: ``\textit{Left me feeling like couch potato. Uh, I don't understand what that means. So I'll just. I'll just be in the middle.}''

\textbf{Interpretation:} ``Recharged'' and ``gave me energy'' were read differently.
\textit{Recharge} presupposes a depleted state and was read as a short restorative break; \textit{energize} was read as similar to, but more general than, motivation.
Several respondents throughout the iterations stated that they think the `couch potato' item is a fun and refreshing item that resonates with their experience.
For example, P1: ``\textit{Hehe, couch potato. I do have the impression that I stay glued to my phone too long (on TikTok).}''

\scale{Inspiration}

\begin{likert}
\likertitem{\textbf{something} that stimulated my creativity}
\likertitem{\textbf{something} that stimulated my imagination}
\likertitem{\textbf{something} that sparked new ideas}
\likertitem{\textbf{something} that inspired me}
\end{likert}

\textbf{Development:} This scale was close to the intended meaning from the first iteration.
One item was dropped: ``something that got me thinking,'' which was interpreted inconsistently; sometimes as intended and other times as a reflection on how the recommendations were generated.

\textbf{Interpretation:} Respondents perceive a fundamental difference between inspiration and creativity.
Creativity was associated to art and hobbies, and carried an activating effect.
For example, P24 states: ``\textit{Stimulated my creativity. I'm gonna disagree strongly 'cause I don't really have any additional creative feelings of, like, building something new or, you know, painting.}''
Inspiration was interpreted more broadly, extending to domains like studying, career, and your social network as P21 described: ``\textit{if it's something about school or careers or even family, it also inspires me.}''
P23 noted a difficulty in answering this scale since whether content stimulated creativity or inspiration may only reveal itself in the future.
Lastly, P3 highlights the importance of specifying the items more concretely based on the goal of the study: ``\textit{inspired me in what sense?}''
Hence, it should be specified what the goal is, as in ``inspired me to cook healthy'' or ``sparked new ideas to save energy.''

\scale{Impact}

\begin{likert}
\likertitem{\textbf{something} that might improve my life}
\likertitem{\textbf{something} that might have a positive impact on my life}
\end{likert}

\begin{likert}
\likertitem{\textbf{something} harmful to me}
\likertitem{\textbf{something} toxic to me}
\end{likert}

\begin{likert}
\likertitem{\textbf{mainly} content that made me feel like I was wasting my time}
\end{likert}

\textbf{Development:} Negative items using \textit{obstruct} or \textit{obstacle} were dropped: they were not universally understood and were perceived as excessively strong, as P7 put it: ``\textit{An obstacle to improve your life is maybe a bit too dramatic.}''

``Something that might have a negative impact on my life'' was dropped in the final iteration because respondents also counted bad news under it.
For example P24: ``\textit{I encountered something that might have a negative impact on my life. Well, there was a lot of US news and yeah. I get paid in US dollars.}''
The interpretation makes sense.
However, the intended meaning is whether receiving the recommendation might have a positive impact, and knowing bad news might actually be helpful.
Hence, valence (is it good or bad news) and value (is it good or bad to know this) are confused.
Therefore, we added the item ``\textbf{something} toxic to me'' which moved from the negative dimension of \textit{resonance}.
Respondents read \textit{toxic} as a negative change rather than a negative feeling, which is what this scale targets, whereas negative \textit{resonance} targets a negative emotional reaction.
P19 separated the two: ``\textit{I think toxic indicates an impact of how I'm thinking, and disturbing is more of a state of emotion I feel afterwards for me.}''
\textit{Harmful} was moved here from the negative dimension of \textit{usefulness} for the same reason, since respondents understood it as close to \textit{toxic}.

\textbf{Interpretation:} The two positive items are perceived to have an hierarchical order.
``Might improve my life'' was generally perceived as stronger than ``might have a positive impact on my life.''
For example, P19: ``\textit{I think improving or I associate improving with a change, an immediate change, and a positive impact is\ldots\ it has slightly less impact. It's just an addition, a positive addition to circumstances.}''
P26's interpretation reflects the intended meaning of harmful: ``\textit{Something that might be harmful. I think the creepy girl that I mentioned to you, like she appears in my feed all the time, and I don't know why. I think like she's on this like massive weight loss journey, and that's why she appears to me. And yeah, it's constant clips about this, about what she, what she's doing. But I don't, I don't know if she's doing it in a very healthy way, and I also think she's pushing messages that are a bit harmful in the sense that she says stuff like, `Men like me more now,' or something like that, which is a bit yeah, harmful for certain audiences.}''
Lastly, ``content that made me feel like I was wasting my time'' is interpreted as an opportunity cost of time that could have been better spent in a different way as P19 illustrates: ``\textit{I would be like, `Okay, this is not for me. I'm wasting my time right now. I could do something better like going outside or make myself a cup of tea.'}''

\scale{Empowering}

\begin{likert}
\likertitem{\textbf{something} positive for my self-esteem}
\end{likert}

\begin{likert}
\likertitem{\textbf{something} negative for my self-esteem}
\end{likert}

\textbf{Development:} This scale required most change.
We dropped all the items from the first iteration.
These were explicitly asking about feeling empowered, feeling capable, boosting confidence et cetera.
However, ``empowering'' is not understood by many.
Other items are perceived as vague such as ``content that made me feel capable'' and worded too strongly such as ``boosted my confidence.''
Also in the second iteration most items were dropped since they were perceived as not applicable e.g., ``makes me feel less confident about myself'' or too vague like ``content that makes me believe in my skills.''
The final iteration reduced the scale to the single formulation that respondents perceived both applicable and understandable, namely self-esteem.

\textbf{Interpretation:} All respondents interpreted the final scale as intended, and they associated a range of relevant experiences with them: affirming content and content that supports confidence on the positive side, and imposter syndrome and negative social comparison on the negative side.
P21 describes the latter: ``\textit{they post their achievements, which is of course a good thing, like a normal thing. But sometimes it affects me. Sometimes of course you see like either someone younger or someone your age, at the same age that's like doing so many things already.}''

\scale{Useful}

\begin{likert}
\likertitem{\textbf{something} useful for a problem I was facing}
\likertitem{\textbf{something} helpful}
\likertitem{\textbf{something} that might prove useful someday}
\end{likert}

\begin{likert}
\likertitem{\textbf{something} that might be misleading}
\likertitem{\textit{(\textbf{mainly} content that made me feel like I was wasting my time)}}
\end{likert}

\textbf{Development:} ``Something valuable'' was dropped since it was interpreted as too broad and abstract: it could point to any condition of the enriching component.
The remaining items only focus on perceived usefulness/helpfulness, which means that the content is instrumental to some goal.

\textbf{Interpretation:} The items reflect this narrow instrumental interpretation as, for example, P26 illustrates when she rejects an item: ``\textit{Something useful for a problem I was facing. No. No problem solving. Something helpful. No. Strongly disagree. Something that might prove useful someday. The frog will not be inherently useful. She will just be pretty.}''

``Mainly content that made me feel like I was wasting my time'' appears in both this scale and \textit{impact}, but it takes a different meaning in each.
Here it refers to time wasted while performing a specific task.
This assumes that the user has a specific task for which the platform is being used.
The scale is consequently most informative in task-driven settings, though it remains usable in aimless browsing, where it can capture background serendipity.

Lastly, misleading is understood as content that should be treated with suspicion or cannot be trusted, which matches the intended meaning.
P27: ``\textit{Misleading? No, but I do often get that kind of [ads], like, `you should use this lotion for your skin' which can be super misleading.}''

\scale{Resonance}

\begin{likert}
\likertitem{\textbf{something} that moved me}
\likertitem{\textbf{something} meaningful to me}
\likertitem{\textbf{something} touching}
\end{likert}

\begin{likert}
\likertitem{\textbf{something} annoying}
\likertitem{\textbf{something} disturbing}
\likertitem{\textbf{mainly} content that felt empty}
\end{likert}

\textbf{Development:} ``Something toxic'' was moved to \textit{impact} in the final iteration, since the negative dimension here targets negative affect while toxic was read as a negative change.
``Empty'' and ``superficial'' were both used to ask about a lack of resonance.
Finally, superficial was dropped because it was frequently not understood, for example P25: ``\textit{Superficial? Superficial, is that an existing word?}''

\textbf{Interpretation:} The scale captures two related elements of resonance.
``Touching'' captures an emotional reaction as P19 expresses: ``\textit{And I also think something touching because I really find those kittens cute,}'' while meaningful is interpreted as a positive relation with values of a person as P19 explains: ``\textit{Videos about like woman rights for me are very meaningful.}''
Both elements are related, but not necessarily always co-occur as P20 illustrates when answering somewhat agree on meaningful but somewhat disagree on touching and moving when reflecting about a podcast about Albanian history.

\section{Manual of the Serendiptionnaire}
\label{sec:manual}

The sections below describe how the Serendiptionnaire can be used and should be scored.

\subsection{Modular: Choosing the Scales that Characterize the Flavor(s) of Interest}

The Serendiptionnaire is modular; it is not intended to be administered or scored in full.
The researcher first decides which flavor, or flavors, of experienced serendipity they aim to design for or study.
Once the target flavor is clear, the instrument is used to evaluate whether that flavor occurred.
A flavor is simply a combination of conditions.
For instance, in previous research we defined taste-broadening serendipity as unusual, intriguing, and any of the fortuitous conditions~\citep{binst_let_2026}.
Measuring taste-broadening serendipity therefore requires the unusual scale, the intriguing scale, and all positive scales belonging to fortuitous.
\citet{binst_what_2025} can provide inspiration but researchers are free to label new serendipity flavors.
However, flavors should be clearly defined and \textbf{only the scales that characterize the targeted flavor(s) should be included}.

\subsection{Scoring Experienced Serendipity}

Each scale measures a single condition through multiple items.
The items describe different expressions of that condition, and each expression is sufficient on its own: a respondent who answers affirmative on one but not the others has still met the condition.
This sets these scales apart from reflective scales, whose items are interchangeable indicators of a latent construct and are therefore expected to correlate.
Here, weak inter-item correlation is not a problem, and internal consistency statistics such as Cronbach's alpha do not apply.
This has a direct consequence for scoring: \textbf{a scale is scored by taking the maximum across its items:} the condition is satisfied to the degree that its most strongly realized expression holds, and it counts as present once this maximum reaches ``somewhat agree.''

Two levels of logic then govern the classification.
\textbf{Within a scale, the characteristic is satisfied disjunctively:} it holds when at least one item reaches the threshold, following the maximum rule.
\textbf{Across the scales that define a flavor, the characteristics are combined conjunctively:} the flavor is classified as present only when all of its defining characteristics hold at the same time.
For taste-broadening serendipity, this results in the following formula:
\begin{quote}
Taste-broadening serendipity = (at least one fortuitous condition is fulfilled) $\land$ (at least one unusual item is fulfilled) $\land$ (at least one intriguing item is fulfilled).
\end{quote}

The final score is binary; either the user experienced taste-broadening serendipity or not.

\subsection{Adapting the Items to the Research Context}

The Serendiptionnaire contains those items that best capture the intended meaning of experienced serendipity in an abstract way.
However, as discussed in Section~\ref{sec:intended} experienced serendipity is intertwined with intended and afforded serendipity~\citep{smets_intended_2025}.
Therefore, the intended and afforded levels influence how the instrument should be used.
Together they describe the context to which the items should be specified: depending on which user experience the design intends to stimulate, and how the feature or algorithm affords that experience, the items of the Serendiptionnaire should be adapted.
Two examples to illustrate this adaptation:
\begin{enumerate}
  \item Consider a cooking app intended to inspire healthier eating by nudging users toward healthy recipes.
  A core condition here is inspiring, but it should be adapted to the specific goal: ``something that inspired me'' becomes ``a recipe that inspired me to cook something healthy.''
  \item Consider a design intended to motivate exercise by recommending personalized activities.
  A core condition here is energizing, and it should be measured at the level of the activities: ``something that motivated me'' becomes ``this activity motivates me to exercise.''
\end{enumerate}

Also the level of analysis should be specified.
The conjunctive rule of the previous section states that the defining characteristics of a flavor must hold at the same time, but it leaves open the object against which they are jointly judged.
Fixing that object --- the referent of the measurement --- is therefore essential.
An unfixed referent leads to confussion: a respondent may judge fortuitous with one object in mind and enriching with another, even though experienced serendipity requires both to hold for the same object.
\textbf{A flavor is identified only when its characteristics are satisfied by the same referent.}

\subsubsection{The Recommendation as Referent}

In the context of RecSys, the most intuitive referent is the individual recommendation.
Each relevant item is evaluated per recommendation, the conjunctive rule is applied to that recommendation, and every scale is anchored to the same object.
Evaluating many recommendations makes the questionnaire repetitive.
A flowchart design might reduce the burden: ask the most discriminative scale first, then ask the second most discriminative scale only of the items that passed the first filter.
Every filter must be a core characteristic of the flavor.

\begin{quote}
\textit{Example:} Consider a book recommender designed to surface books that resonate emotionally because they are related to struggles the users currently face.
Resonance is then a core condition of the serendipity flavor we target, and the first question after each recommendation should address it: whether the book seems moving, meaningful, or touching.
Books scoring low on resonance cannot qualify as resonating serendipity; so, if a book scores negative, the remaining scales can be skipped for that book.
\end{quote}

\subsubsection{The Session as Referent}

A flavor may instead be conceptualized as a property of the session as a whole.
For example, the combination or the trajectory of the recommendations.
The session then becomes the referent.
However, it is important to pilot whether the respondents do really think about the session as a whole in the study's context since the session is a more diffuse object to hold in mind and evaluate than the single recommended items.

\begin{quote}
\textit{Example:} \citet{taramigkou_escape_2013} developed an algorithm that guides users into a target genre by recommending a path of songs leading from their established preferences toward that genre.
Since the intended serendipity lies in warming listeners to the genre rather than in any single song, taste-broadening serendipity is probably best measured at the session level.
\end{quote}

\section{Discussion}

Existing serendipity operationalizations operate at different levels of serendipity.
Furthermore, the operationalizations measuring experienced serendipity are fragmented.
This fragmentation obstructs comparison between studies, synthesis of findings, and progress in research on serendipity in RecSys~\citep{binst_how_2024, busch_towards_2022, kotkov_challenges_2016}.
\citet{binst_what_2025} contributed to solving this issue by developing a conceptual framework that reconciled the conceptual ambiguity of previous research by allowing for different flavors of experienced serendipity.
However, a validated questionnaire measuring experienced serendipity was still missing.
This paper fills this gap in the literature.

Serendipity has been characterized as a mysterious and slippery concept that is very difficult to capture~\citep{makri_making_2014}.
Especially its subjective nature makes it hard to operationalize experienced serendipity~\citep{kang_exploring_2025, tokutake_universal_2025}.
To operationalize this subjective and complicated concept, we conducted twenty-seven cognitive interviews to formulate a questionnaire that is interpreted as intended following~\citep{binst_what_2025}.
Based on this study we propose some small updates to our original conceptualization of experienced serendipity (see Section~\ref{sec:theory}).
However, first we discuss some important empirical implications of this study.

\subsection{Empirical Implications: a Ground Truth for Validating Offline Metrics of Serendipity}

RecSys research often evaluates serendipity using system-centered metrics~\citep{binst_how_2024, kang_exploring_2025, kotkov_survey_2016}.
These metrics calculate scores based on item features, interaction histories, and popularity data~\citep{murakami_metrics_2008, perner_unexpectedness-augmented_2015, kotkov_investigating_2018}.
In doing so, they substitute researcher assumptions for actual user experiences~\citep{chen_how_2019, kotkov_challenges_2016}.
However, researchers rarely validate whether these system-centered metrics actually reflect user perception~\citep{chen_user_2017}.
This is problematic since we need validated measurement instruments for progressing research~\citep{jannach_escaping_2020, kang_exploring_2025}.

Where validation has been attempted, the results are discouraging.
\citet{chen_user_2017} correlated system-centered serendipity metrics with a single user-centered item (``the item recommended to me is a pleasant surprise'') and found correlations that were small and largely nonsignificant.
We can understand these results in two ways: Either the metrics are invalid or the construct they operationalize is underspecified.
A single item asking about pleasant surprises simplifies experienced serendipity into one undifferentiated judgment.
In contrast, the conceptual framework underlying the Serendiptionnaire recognizes that serendipity comes in distinct flavors~\citep{binst_what_2025}.
Hence, instead of validating a generic serendipity metric, researchers should focus on metrics that operationalize either specific flavors or dimensions of experienced serendipity.
The Serendiptionnaire is a validated and nuanced questionnaire that can be used for this.

Take, for instance, three of the most frequently used system-centered operationalizations of serendipity.
\citet{ge_beyond_2010} operationalize serendipity as a system centered metric combining relevance and unexpectedness, where unexpectedness is operationalized as those items that would not be recommended by a primitive recommender system (which is often a recommender system based on popularity).
Following our framework, this might be predictive for relevant-and-difficult-to-find serendipity, and should therefore be validated using the scales measuring useful and difficult-to-find encounters.
Conversely, \citet{ricci_evaluating_2011} operationalized serendipitous recommendations through its distance from items already in the user profile, rewarding for example books written by writers the reader had read least often.
This metric might be predictive for taste-broadening serendipity and could therefore be validated through the unusualness and intriguing scales.
Finally, \citet{kaminskas_measuring_2014} equate serendipity with surprise, which they operationalize through the pointwise mutual information between a candidate item and the items in the user profile.
This metric claims predictive power for the surprise condition of experienced serendipity, and should therefore be validated using the surprise scale.
Hence, the Serendiptionnaire advances the field by providing a stable, empirical foundation for future research:
\begin{itemize}
  \item \textbf{It supports metric development.}
  Building system-centered metrics becomes a targeted problem.
  Instead of guessing how ``unexpectedness'' translates into a system-centered metric, researchers can empirically test which system-centered metrics (like the primitive recommender baseline proposed by \citet{ge_beyond_2010}) is predictive for unexpectedness as measured through the Serendiptionnaire.
  \item \textbf{It enables meaningful comparison across studies.}
  Researchers have operationalized serendipity in fragmented ways~\citep{binst_how_2024, kotkov_dark_2024, ziarani_serendipity_2021}.
  The Serendiptionnaire standardizes the vocabulary, allowing researchers to synthesize results and build on past work rather than talking past one another.
  \item \textbf{It provides ground truth for LLM evaluation.}
  As researchers increasingly employ LLMs to evaluate serendipity in RecSys~\citep{fu_art_2024, kang_exploring_2025, tokutake_universal_2025}, the Serendiptionnaire can offer a rigorous benchmark to evaluate their performance.
  Furthermore, the conceptual framework and specific items in the Serendiptionnaire can be used as context when prompting an LLM for evaluating recommendations for a specific flavor of experienced serendipity.
  This might improve LLMs performance in both evaluating and stimulating experienced serendipity in RecSys.
  \item \textbf{It provides insight for improvement.}
  When an intervention fails to increase serendipity, the Serendiptionnaire will provide insight into why.
  If an algorithm increases the ``unusualness'' of recommendations but fails to trigger the ``intriguing'' dimension, researchers know they must focus on stimulating intrigue, not just novelty.
  Hence, the Serendiptionnaire provides granular feedback, previously unavailable, for designers to refine their RecSys.
\end{itemize}

\subsection{Theoretical Contribution: Updates to Our Conceptualization of Experienced Serendipity}
\label{sec:theory}

\paragraph{Unintentionality is no longer required.}
The original conceptualization stated that a serendipitous experience must be unintentional and additionally satisfy one further fortuitous condition.
However, conceptualizing unintentional as essential proved unproductive; user sessions are often intentional on some level, even if the user intention is vague.
\citet{binst_let_2026}, for instance, deployed a music discovery application designed for exploration outside a listener's typical preferences.
There is a clear user goal when using this app, so it is difficult for users to state that they encountered something they were not looking for.
Nevertheless, if a user discovers a song that broadens their taste, we would still call this taste-broadening serendipity, even though the user might not perceive the discovery as unintentional.
Therefore, we update this component to be like the other components, meaning that at least one condition should be fulfilled, and unintentional is no longer necessary but just one of the potential conditions to be fulfilled.

\paragraph{Uncovering is defined more broadly.}
Uncovering was defined as revealing a latent interest.
However, also this conceptualization proved to be unproductively strict.
It is more productive to define uncovering as discovering something from a new perspective.
This can entail revealing a latent interest, but also realizing something new about, for example, a song you already liked.
The broader definition implies that uncovering is a condition of refreshing as well as fortuitous because it describes both a manner of encountering, it is discovered from a new perspective, and a relationship between the user and the item, the content is encountered through a fresh perspective.

\paragraph{Experienced serendipity can score negatively on enriching.}
In \citet{binst_what_2025}, we state that experienced serendipity cannot score negatively on the enriching scale.
However, we also noticed that this rule is too strict.
User experiences are very complicated and can contain both negative and positive elements at the same time.
For example, encountering negative news can be very useful for the user, helping them to mitigate the negative consequences.
Nevertheless, negative news will probably be perceived as disturbing, but this does not make it a negative experience.
Therefore, we drop this rule from the conceptual framework.
This results in the updated framework presented in Table~\ref{tab:framework}.

\begin{table}[htbp]
  \caption{Schematic overview of our conceptualization of experienced serendipity. Experienced serendipity has three main components: fortuitous, refreshing and enriching. These main components can be fulfilled through different conditions. Uncovering is a cross-cutting condition that belongs to both the fortuitous and refreshing component.}
  \label{tab:framework}
  \small
  \setlength{\tabcolsep}{5pt}
  \begin{adjustbox}{max width=\linewidth, center}
  \begin{tabular}{@{}l|l|l|l@{}}
    \toprule
    \textbf{Main Component} & \textbf{Condition} & \textbf{Expression} & \textbf{Opposite} \\
    \midrule
    \multirow{4}{*}{\begin{tabular}[c]{@{}l@{}}\textit{\textbf{Fortuitous vs.}}\\\textit{\textbf{Anticipated}}\end{tabular}}
      & Unintentional & Encountered & Explicitly searched \\
      & Difficult-to-find & RecSys facilitated discovery & Easily discovered without RecSys \\
      & Unexpected & Surprised & Unsurprising \\
    \cmidrule{2-4}
      & \multirow{2}{*}{Uncovering} & \multirow{2}{*}{Discovering something from a new perspective} & \multirow{2}{*}{Fixed perspective} \\
    \cmidrule{1-1}
    \multirow{4}{*}{\begin{tabular}[c]{@{}l@{}}\textit{\textbf{Refreshing vs.}}\\\textit{\textbf{Boring}}\end{tabular}}
      & & & \\
    \cmidrule{2-4}
      & Novel & Completely unknown & Consumed \\
      & Unusual & Different compared to regular consumption & Hyperfixation \\
      & Taste-reincarnation & Reconnecting & Weakening connection \\
    \midrule
    \multirow{7}{*}{\begin{tabular}[c]{@{}l@{}}\textit{\textbf{Enriching vs.}}\\\textit{\textbf{Brainrot}}\end{tabular}}
      & Intriguing & Sparks curiosity & Uninteresting \\
      & Energizing & Motivating & Demotivating \\
      & Inspiring & Sparks creativity & Stifles creativity \\
      & Impact & Positive influence on life & Negative influence \\
      & Empowering & Empowering & Disempowering \\
      & Usefulness & Useful & Misleading \\
      & Resonance & Touching & Out-of-touch \\
    \bottomrule
  \end{tabular}
  \end{adjustbox}
\end{table}




\subsection{Limitations and Future Studies}

\paragraph{Ease of use.}
The Serendiptionnaire cannot be copied and pasted; you have to adapt it to your use case.
We made the Serendiptionnaire intentionally flexible.
Because every scale starts simply with ``I encountered'' and most items begin with ``something,'' researchers can easily tailor the phrasing to fit their research context.
As explained in Section~\ref{sec:intended} the Serendiptionnaire is embedded within a theoretical framework that distinguishes between different levels of serendipity that are intertwined.
Researchers are encouraged to use this framework (see Section~\ref{sec:manual}) to pick the right conditions that characterize the flavor you intend to study, and to adapt the items to the specific context in which this flavor will be studied.
These specified items should also be piloted before actual use.

\paragraph{Exhaustiveness.}
Furthermore, we do not think the Serendiptionnaire is currently exhaustive; we believe there are more conditions to add.
For example, we would add synchronicity, i.e., a meaningful coincidence~\citep{bogers_micro-serendipity_2013}, to the fortuitous component.
Similarly, we think there are additional conditions appropriate to capture the nuances of an enriching user session beyond those currently included in the framework.

\paragraph{Generalizability.}
Lastly, during this study, we developed the Serendiptionnaire for evaluating users' experience interacting with RecSys.
However, it would be interesting to study whether the Serendiptionnaire can also be used to evaluate experienced serendipity in the context of other digital systems.

\section{Conclusion}

Experienced serendipity is subjective and slippery, and it expresses itself in many flavors.
These features make it difficult to develop a validated instrument to measure experienced serendipity.
Through cognitive interviews, a methodology developed to formulate questionnaire items that are well-understood, we create the Serendiptionnaire; a questionnaire that operationalizes experienced serendipity as conceptualized in our earlier work.
We conducted twenty-seven cognitive interviews spread over three iterations to refine the Serendiptionnaire until respondents of different ages, education levels, and native languages understand its items as intended.

The Serendiptionnaire contributes to the literature in several ways.
It turns system-centered metric development for serendipity into a targeted empirical problem; by using the metric as a predictor for the relevant scale(s) of the Serendiptionnaire, the metric can be validated.
Additionally, the Serendiptionnaire can also work as a benchmark for comparing LLM evaluations of serendipitous recommendations.
Furthermore, it can also be used as context when prompting an LLM to evaluate a specific flavor.
Finally, the Serendiptionnaire provides fine-grained insight into why a specific design succeeds or fails to stimulate experienced serendipity.

\section{Ethical and Human Subjects Considerations}
The study protocol was reviewed and approved by the Ethics Committee of our institute.
Informed consent was obtained from all participants prior to the interviews.
Participants were explicitly informed of their right to withdraw from the study at any time without penalty.

\begin{acks}
This work was supported by the Research Foundation Flanders (FWO)  under grant number S006323N.
We are grateful to all our participants for generously contributing their time and insights to this study.
We further thank our colleagues, in particular Lien Michiels and the members of the reading group at SMIT for their thorough reviews of the manuscript and the many constructive suggestions.
\end{acks}

\bibliographystyle{ACM-Reference-Format}
\bibliography{bibliography}

\appendix

\clearpage
\section{The Full Serendiptionnaire}
\label{app:serendiptionnaire}

\noindent This is the final version of the Serendiptionnaire, as described in Section~\ref{sec:serendiptionnaire}.

\subsection*{Introduction}

\noindent\textit{In this questionnaire, we are interested in how you feel about the content you encountered \textbf{during the session you just finished}.
There are no right or wrong answers, so please don't overthink the questions, just answer based on what comes to mind first.}

\subsection*{Fortuitous vs.\ Anticipated}

\scale{Unintended}

\begin{likert}
\likertitem{\textbf{something} I did not specifically search for during this session}
\likertitem{\textbf{something} I was not specifically looking for during the session}
\end{likert}

\scale{Difficult to Find}

\begin{likert}
\likertitem{a hidden gem (\textbf{something} interesting but that I did not know yet, or overlooked until now)}
\likertitem{\textbf{something} that might have been more difficult to discover without the help of this platform}
\end{likert}

\begin{likert}
\likertitem{\textbf{mainly} content that I would have easily discovered without the platform}
\end{likert}

\scale{Unexpected}

\begin{likert}
\likertitem{\textbf{something} that surprised me}
\likertitem{\textbf{something} unexpected}
\end{likert}

\begin{likert}
\likertitem{\textbf{nothing} surprising to me}
\likertitem{\textbf{only} content that I expected to find}
\end{likert}

\scale{Uncovering}

\begin{likert}[this scale cross-cuts the fortuitous and refreshing component.]
\likertitem{\textbf{something} that helped me realize something new}
\likertitem{\textbf{something} that I did not know could be useful to me}
\likertitem{\textbf{something} that made me aware about something I did not know about before}
\likertitem{\textbf{something} that made me notice details I did not notice before}
\end{likert}

\subsection*{Refreshing vs.\ Boring}

\scale{Novelty}

\begin{likert}
\likertitem{\textbf{at least one item} I had not seen before}
\likertitem{\textbf{at least one item} I did not know about yet}
\end{likert}

\begin{likert}
\likertitem{\textbf{mainly} items I had seen before}
\likertitem{\textbf{mainly} items I knew about already}
\end{likert}

\scale{Unusual}

\begin{likert}
\likertitem{\textbf{something} different from the type of content I normally consume}
\likertitem{\textbf{something} different from my typical preferences}
\end{likert}

\begin{likert}
\likertitem{\textbf{mainly} the same genre of content I normally consume}
\end{likert}

\scale{Taste-Reincarnation}

\begin{likert}
\likertitem{\textbf{something} that feels nostalgic to me}
\likertitem{\textbf{something} that reconnected me to something that I previously liked}
\end{likert}

\begin{likert}
\likertitem{\textbf{something} from my personal past that I prefer to avoid}
\likertitem{\textbf{something} that reduced my interest in something I previously liked}
\end{likert}

\subsection*{Enriching vs.\ Brainrot}

\scale{Intrigue}

\begin{likert}
\likertitem{\textbf{something} that sparked my interest}
\likertitem{\textbf{something} that stimulated my curiosity}
\end{likert}

\begin{likert}
\likertitem{\textbf{mainly} content that failed to grab my attention}
\end{likert}

\scale{Energy}

\begin{likert}
\likertitem{\textbf{something} that recharged me}
\likertitem{\textbf{something} that gave me energy}
\likertitem{\textbf{something} that motivates me to \ldots}
\end{likert}

\begin{likert}
\likertitem{\textbf{something} that lowered my energy}
\likertitem{\textbf{something} that left me feeling like a couch potato (lazy and without energy).}
\end{likert}

\scale{Inspiration}

\begin{likert}
\likertitem{\textbf{something} that stimulated my creativity}
\likertitem{\textbf{something} that stimulated my imagination}
\likertitem{\textbf{something} that sparked new ideas}
\likertitem{\textbf{something} that inspired me}
\end{likert}

\scale{Impact}

\begin{likert}
\likertitem{\textbf{something} that might improve my life}
\likertitem{\textbf{something} that might have a positive impact on my life}
\end{likert}

\begin{likert}
\likertitem{\textbf{something} harmful to me}
\likertitem{\textbf{something} toxic to me}
\end{likert}

\begin{likert}
\likertitem{\textbf{mainly} content that made me feel like I was wasting my time}
\end{likert}

\scale{Empowering}

\begin{likert}
\likertitem{\textbf{something} positive for my self-esteem}
\end{likert}

\begin{likert}
\likertitem{\textbf{something} negative for my self-esteem}
\end{likert}

\scale{Useful}

\begin{likert}
\likertitem{\textbf{something} useful for a problem I was facing}
\likertitem{\textbf{something} helpful}
\likertitem{\textbf{something} that might prove useful someday}
\end{likert}

\begin{likert}
\likertitem{\textbf{something} that might be misleading}
\likertitem{\textit{(\textbf{mainly} content that made me feel like I was wasting my time)}}
\end{likert}

\scale{Resonance}

\begin{likert}
\likertitem{\textbf{something} that moved me}
\likertitem{\textbf{something} meaningful to me}
\likertitem{\textbf{something} touching}
\end{likert}

\begin{likert}
\likertitem{\textbf{something} annoying}
\likertitem{\textbf{something} disturbing}
\likertitem{\textbf{mainly} content that felt empty}
\end{likert}

\clearpage
\section{Iteration 1}
\label{app:iteration1}

\noindent This is the questionnaire tested in the first of the three iterations through which the Serendiptionnaire was tested and refined.

\noindent The first version had no introduction.

\subsection*{Fortuitous vs.\ Anticipated}

\scale{Unintended}

\begin{bipolar}{During my session on the platform}
\bipolaritem{I only encountered content \textbf{I explicitly searched}}{I encountered content \textbf{I did not plan to find}}
\bipolaritem{I only encountered content \textbf{intentionally}}{I encountered content \textbf{unintentionally}}
\bipolaritem{I only encountered content \textbf{on purpose}}{I encountered content \textbf{not on purpose}}
\bipolaritem{I only encountered content \textbf{deliberately}}{I encountered content \textbf{accidentally}}
\end{bipolar}

\scale{Difficult to Find}

\begin{likertq}{During my session on the platform}
\likertitem{I encountered a \textbf{hidden gem}}
\end{likertq}

\begin{bipolar}{During my session on the platform}
\bipolaritem{I only encountered content that would be \textbf{easy to find on my own}}{I encountered content that would be \textbf{difficult to find on my own}}
\bipolaritem{I could have found this content \textbf{easily on my own}}{The platform helped me discover content that would be \textbf{difficult to find on my own}}
\end{bipolar}

\scale{Unexpected}

\begin{bipolar}{My session on the platform}
\bipolaritem{Felt \textbf{predictable}}{\textbf{Surprised} me}
\bipolaritem{Was \textbf{unsurprising}}{Took me by \textbf{surprise}}
\bipolaritem{Was as \textbf{expected}}{Was \textbf{unexpected}}
\end{bipolar}

\begin{bipolar}{During my session on the platform}
\bipolaritem{I encountered \textbf{predictable} content}{I encountered \textbf{random} content}
\bipolaritem{The content I encountered \textbf{did not surprise me}}{I encountered \textbf{surprising} content}
\bipolaritem{I only encountered \textbf{expected} content}{I encountered \textbf{unexpected} content}
\end{bipolar}

\scale{Uncovering}

\begin{bipolar}{My session on the platform}
\bipolaritem{\textbf{Reinforced what I already knew about myself}}{Made me \textbf{realize something new about myself}}
\bipolaritem{\textbf{Confirmed what I already believed}}{Helped me \textbf{discover something new about myself}}
\end{bipolar}

\begin{likertq}{My session on the platform}
\likertitem{Lead me to \textbf{new self-understanding}}
\likertitem{Helped me \textbf{uncover something I was not aware of}}
\end{likertq}

\begin{likertq}{During my session on the platform}
\likertitem{I encountered content \textbf{I didn't know I was looking for}}
\likertitem{I encountered something \textbf{I didn't know I needed}}
\likertitem{I \textbf{realized something new about myself}}
\likertitem{I had a \textbf{personal revelation}}
\end{likertq}

\subsection*{Refreshing vs.\ Boring}

\scale{Novelty}

\begin{bipolar}{My session on the platform}
\bipolaritem{\textbf{Introduced nothing new} to me}{\textbf{Introduced something new} to me}
\end{bipolar}

\begin{bipolar}{During my session on the platform}
\bipolaritem{I only encountered content that \textbf{I already knew}}{I encountered \textbf{novel} content}
\bipolaritem{I only encountered \textbf{familiar} content}{I encountered \textbf{unfamiliar}}
\bipolaritem{I only encountered content that \textbf{I had seen before}}{I encountered content \textbf{I hadn't seen before}}
\bipolaritem{I only encountered content \textbf{I already consumed}}{I encountered content \textbf{I didn't consume yet}}
\end{bipolar}

\begin{likertq}{My session on the platform}
\likertitem{Helped me discover novel content}
\end{likertq}

\scale{Unusual}

\begin{bipolar}{My session on the platform}
\bipolaritem{Felt \textbf{constraining}}{\textbf{Broadened my horizon}}
\bipolaritem{Felt \textbf{confining}}{\textbf{Expanded my perspective}}
\end{bipolar}

\begin{bipolar}{During my session on the platform}
\bipolaritem{I encountered \textbf{boring} content}{I encountered \textbf{refreshing} content}
\bipolaritem{I encountered \textbf{the same content I normally watch}}{I encountered content that \textbf{differs from what I normally watch}}
\bipolaritem{I encountered content that \textbf{matched my usual preferences}}{I encountered content that differs from \textbf{my usual choices}}
\bipolaritem{I encountered content that \textbf{aligned with my usual viewing habits}}{I encountered content \textbf{outside my usual viewing habits}}
\bipolaritem{I encountered content that \textbf{mirrored my usual consumption pattern}}{I encountered content that \textbf{broke from my usual consumption pattern}}
\bipolaritem{I encountered content that \textbf{fit my typical interests}}{I encountered content \textbf{outside my typical interests}}
\end{bipolar}

\scale{Taste-Reincarnation}

\begin{bipolar}{My session on the platform}
\bipolaritem{Made me encounter \textbf{content from the past I rather avoid}}{Made me feel \textbf{nostalgic}}
\bipolaritem{\textbf{Reduced my interest in things I used to enjoy}}{\textbf{Reconnected me with the past}}
\bipolaritem{Made me \textbf{lose interest in what I once liked}}{Helped me \textbf{rediscover my past self}}
\bipolaritem{Made my \textbf{previous passions feel less exciting}}{\textbf{Revived past passions}}
\end{bipolar}

\subsection*{Enriching vs.\ Brainrot}

\scale{Intrigue}

\begin{bipolar}{My session on the platform}
\bipolaritem{Left me \textbf{uninterested}}{Triggered my \textbf{curiosity}}
\bipolaritem{\textbf{Turned me off}}{\textbf{Intrigued me}}
\bipolaritem{Left me \textbf{indifferent}}{\textbf{Fascinated} me}
\end{bipolar}

\begin{bipolar}{During my session on the platform}
\bipolaritem{I only encountered content that \textbf{left me uninterested}}{I encountered \textbf{intriguing content}}
\bipolaritem{I only encountered content that \textbf{failed to grab my attention}}{I encountered content that \textbf{triggered my curiosity}}
\bipolaritem{I only encountered content that made me \textbf{feel indifferent}}{I encountered content that \textbf{fascinated me}}
\end{bipolar}

\scale{Energy}

\begin{bipolar}{My session on the platform}
\bipolaritem{\textbf{Drained my energy}}{\textbf{Energized} me}
\bipolaritem{\textbf{Demotivated} me}{\textbf{Motivated} me}
\bipolaritem{Left me feeling like a \textbf{couch potato}}{\textbf{Recharged} me}
\bipolaritem{\textbf{Drained my energy}}{\textbf{Gave me energy}}
\end{bipolar}

\begin{bipolar}{During my session on the platform}
\bipolaritem{I encountered content that \textbf{drained my energy}}{I encountered content that \textbf{energized} me}
\bipolaritem{I encountered content that \textbf{demotivated} me}{I encountered content that \textbf{motivated} me}
\end{bipolar}

\scale{Inspiration}

\begin{likertq}{During my session on the platform}
\likertitem{I encountered content that \textbf{sparked new ideas}}
\end{likertq}

\begin{likertq}{My session on the platform}
\likertitem{Sparked new ideas}
\likertitem{Inspired me}
\end{likertq}

\begin{bipolar}{During my session on the platform}
\bipolaritem{I encountered content that \textbf{didn't stimulate my imagination}}{I encountered content that \textbf{stimulated my imagination}}
\bipolaritem{I encountered content that left me \textbf{uninspired}}{I encountered content that \textbf{sparked my creativity}}
\end{bipolar}

\begin{bipolar}{My session on the platform}
\bipolaritem{\textbf{Did not stimulate my imagination}}{\textbf{Stimulated my imagination}}
\bipolaritem{Left me \textbf{uninspired}}{\textbf{Sparked my creativity}}
\end{bipolar}

\scale{Impact}

\begin{bipolar}{During my session on the platform}
\bipolaritem{I encountered content that will have a \textbf{negative impact on my life}}{I encountered content that will have a \textbf{positive impact on my life}}
\end{bipolar}

\begin{bipolar}{My session on the platform}
\bipolaritem{Will have a \textbf{negative impact on my life}}{Will have a \textbf{positive impact on my life}}
\bipolaritem{Feels as an \textbf{obstacle to improve my life}}{Might help me \textbf{improve my life}}
\bipolaritem{\textbf{Obstructs personal growth}}{Might stimulate \textbf{personal growth}}
\end{bipolar}

\scale{Empowering}

\begin{bipolar}{During my session on the platform}
\bipolaritem{I encountered content that \textbf{disempowered} me}{I encountered content that \textbf{empowered} me}
\bipolaritem{I encountered content that \textbf{drained my confidence}}{I encountered content that \textbf{boosted my confidence}}
\bipolaritem{I encountered content that made me \textbf{feel incapable}}{I encountered content that made me \textbf{feel capable}}
\bipolaritem{I encountered content that \textbf{diminished my sense of control}}{I encountered content that gave me a \textbf{sense of control}}
\bipolaritem{I encountered content that made me \textbf{doubt myself}}{I encountered content that made me \textbf{believe in myself}}
\end{bipolar}

\begin{bipolar}{My session on the platform}
\bipolaritem{\textbf{Disempowered} me}{\textbf{Empowered} me}
\bipolaritem{\textbf{Drained my confidence}}{\textbf{Boosted my confidence}}
\bipolaritem{Made me \textbf{feel incapable}}{Made me \textbf{feel capable}}
\bipolaritem{\textbf{Diminished my sense of control}}{Gave me a \textbf{sense of control}}
\bipolaritem{Made me \textbf{doubt myself}}{Made me \textbf{believe in myself}}
\end{bipolar}

\scale{Useful}

\begin{bipolar}{During my session on the platform}
\bipolaritem{I encountered content that is \textbf{harmful} to me}{I encountered content that is \textbf{useful} to me}
\bipolaritem{I encountered content that is \textbf{toxic}}{I encountered content that is \textbf{helpful} to me}
\bipolaritem{I encountered content that is \textbf{misleading}}{I encountered content that is \textbf{valuable} to me}
\end{bipolar}

\begin{bipolar}{My session on the platform}
\bipolaritem{Felt \textbf{harmful}}{Felt \textbf{useful}}
\bipolaritem{Felt \textbf{toxic}}{Felt \textbf{helpful}}
\bipolaritem{Felt \textbf{misleading}}{Felt \textbf{valuable}}
\end{bipolar}

\scale{Resonance}

\begin{bipolar}{My session on the platform}
\bipolaritem{Felt \textbf{out-of-touch}}{\textbf{Touched} me}
\bipolaritem{Felt \textbf{alienating}}{\textbf{Resonated} with me}
\bipolaritem{Felt \textbf{empty}}{Felt \textbf{meaningful}}
\bipolaritem{Left me \textbf{indifferent}}{\textbf{Moved} me}
\end{bipolar}

\begin{bipolar}{During my session on the platform}
\bipolaritem{I encountered content that felt \textbf{out-of-touch}}{I encountered content that \textbf{touched} me}
\bipolaritem{I encountered content that felt \textbf{alienating}}{I encountered content that \textbf{raised my spirit}}
\bipolaritem{I encountered content I \textbf{could not relate with}}{I encountered content that \textbf{resonated} with me}
\bipolaritem{I encountered content that \textbf{left me indifferent}}{I encountered content that \textbf{moved} me}
\end{bipolar}

\clearpage
\section{Iteration 2}
\label{app:iteration2}

\noindent This is the questionnaire tested in the second of the three iterations through which the Serendiptionnaire was tested and refined.

\subsection*{Introduction}

\noindent\textit{In this questionnaire, we are interested in how you feel about the content you encountered during the session you just finished.
There are no right or wrong answers so please don't overthink the questions, just answer using your gut feeling.}

\subsection*{Fortuitous vs.\ Anticipated}

\scale{Unintended}

\begin{likertq}{This session on the platform helped me discover}
\likertitem{something \textbf{unintentionally}}
\likertitem{something \textbf{I did not plan to find}}
\likertitem{something \textbf{accidentally}}
\end{likertq}

\scale{Difficult to Find}

\begin{likertq}{This session on the platform helped me discover}
\likertitem{a \textbf{hidden gem}}
\likertitem{something that would be \textbf{difficult-to-find without the help of this platform}}
\end{likertq}

\begin{likertq}{During this session, the platform only showed me}
\likertitem{content that \textbf{I would have easily found without the platform}}
\end{likertq}

\scale{Unexpected}

\begin{likertq}{During this session, the platform only showed me}
\likertitem{the content \textbf{I expected to find}}
\likertitem{content \textbf{that did not surprise me}}
\end{likertq}

\begin{likertq}{This session on the platform helped me discover}
\likertitem{something \textbf{unexpected}}
\likertitem{something \textbf{surprising}}
\end{likertq}

\scale{Uncovering}

\begin{likertq}{This session on the platform helped me discover}
\likertitem{content that \textbf{helped me realize something new about my life}}
\likertitem{something \textbf{I didn't know I was interested in}}
\end{likertq}

\begin{likertq}{This session on the platform helped me discover}
\likertitem{something \textbf{I did not know I could be useful to me}}
\likertitem{something \textbf{I did not know I was looking for}}
\end{likertq}

\subsection*{Refreshing vs.\ Boring}

\scale{Novelty}

\begin{likertq}{This session on the platform helped me discover}
\likertitem{something \textbf{new to me}}
\likertitem{something \textbf{I had not seen before}}
\likertitem{something \textbf{I did not know about yet}}
\end{likertq}

\begin{likertq}{During this session, the platform showed me}
\likertitem{\textbf{nothing new}}
\likertitem{\textbf{nothing I had not seen before}}
\likertitem{\textbf{nothing I did not know about yet}}
\end{likertq}

\scale{Unusual}

\begin{likertq}{During this session on the platform}
\likertitem{I encountered \textbf{the same content I normally watch}}
\likertitem{I felt \textbf{limited in the type of content I encountered}}
\likertitem{I felt \textbf{restricted in the type of content I encountered}}
\end{likertq}

\begin{likertq}{This session on the platform helped me discover}
\likertitem{content that \textbf{differs from what I normally watch}}
\likertitem{content \textbf{outside my typical interests}}
\likertitem{discover content that \textbf{differs from my usual preferences}}
\likertitem{content that \textbf{expanded my perspective}}
\likertitem{content that \textbf{broadened my horizon}}
\end{likertq}

\scale{Taste-Reincarnation}

\begin{likertq}{This session on the platform helped me discover}
\likertitem{something \textbf{nostalgic}}
\likertitem{something that \textbf{reconnected me to the past}}
\end{likertq}

\begin{likertq}{During this session, the platform showed me}
\likertitem{something \textbf{from the past that I would prefer to avoid}}
\likertitem{something that \textbf{reduced my interest in something I once liked}}
\end{likertq}

\subsection*{Enriching vs.\ Brainrot}

\scale{Intrigue}

\begin{likertq}{During this session, the platform only showed me content that}
\likertitem{\textbf{left me indifferent}}
\likertitem{\textbf{failed to grab my attention}}
\end{likertq}

\begin{likertq}{This session on the platform helped me discover}
\likertitem{something that \textbf{sparked my interest}}
\likertitem{something that \textbf{stimulated my curiosity}}
\likertitem{something that \textbf{intrigued me}}
\end{likertq}

\scale{Energy}

\begin{likertq}{This session on the platform helped me discover}
\likertitem{something that \textbf{motivated me}}
\likertitem{something that \textbf{recharged me}}
\likertitem{something that \textbf{energized me}}
\end{likertq}

\begin{likertq}{During my session, the platform}
\likertitem{showed me something that \textbf{demotivated me}}
\likertitem{showed me something that \textbf{drained my energy}}
\likertitem{left me \textbf{feeling like a couch potato}}
\end{likertq}

\scale{Inspiration}

\begin{likertq}{This session on the platform helped me discover}
\likertitem{something that \textbf{got me thinking}}
\likertitem{something that \textbf{stimulated my creativity}}
\likertitem{something that \textbf{stimulated my imagination}}
\likertitem{something that \textbf{sparked new ideas}}
\likertitem{something that \textbf{inspired me}}
\end{likertq}

\scale{Impact}

\begin{likertq}{This session on the platform helped me discover}
\likertitem{something that might \textbf{improve my life}}
\likertitem{something that might have \textbf{a positive impact on my life}}
\end{likertq}

\begin{likertq}{This session, the platform showed me content that}
\likertitem{made me feel like I am \textbf{wasting my time}}
\likertitem{might have \textbf{a negative impact on my life}}
\end{likertq}

\scale{Empowering}

\begin{likertq}{This session, the platform showed me content that}
\likertitem{makes me \textbf{doubt myself}}
\likertitem{makes me feel \textbf{less confident about myself}}
\end{likertq}

\begin{likertq}{This session on the platform helped me discover}
\likertitem{content that \textbf{makes me feel confident}}
\likertitem{content that makes me \textbf{believe in my skills}}
\end{likertq}

\scale{Useful}

\begin{likertq}{This session on the platform helped me discover}
\likertitem{something \textbf{valuable}}
\likertitem{something that is \textbf{useful for a problem I was facing}}
\end{likertq}

\begin{likertq}{This session, the platform showed me}
\likertitem{mainly content that is \textbf{useless}}
\likertitem{content that might be \textbf{misleading}}
\likertitem{content that might be \textbf{harmful}}
\end{likertq}

\scale{Resonance}

\begin{likertq}{This session on the platform helped me discover}
\likertitem{something that \textbf{gave me goosebumps}}
\likertitem{something that \textbf{raised my spirit}}
\likertitem{something that \textbf{moved me}}
\likertitem{something that felt \textbf{meaningful to me}}
\likertitem{something that \textbf{touched me}}
\end{likertq}

\begin{likertq}{This session, the platform showed me}
\likertitem{something \textbf{toxic}}
\likertitem{something that \textbf{disturbed me}}
\likertitem{something that \textbf{annoyed me}}
\likertitem{mainly content that \textbf{felt empty}}
\likertitem{mainly content that \textbf{left me indifferent}}
\end{likertq}

\clearpage
\section{Iteration 3}
\label{app:iteration3}

\noindent This is the questionnaire tested in the third of the three iterations through which the Serendiptionnaire was tested and refined.

\subsection*{Introduction}

\noindent\textit{In this questionnaire, we are interested in how you feel about the content you encountered \textbf{during the session you just finished}.
There are no right or wrong answers so please don't overthink the questions, just answer based on what comes to mind first.}

\subsection*{Fortuitous vs.\ Anticipated}

\scale{Unintended}

\begin{likertq}{I encountered}
\likertitem{\textbf{something} I did not search for}
\likertitem{\textbf{something} I did not plan to find}
\likertitem{\textbf{something} I did not specifically intend to find}
\likertitem{\textbf{something} I was not specifically looking for during the session}
\end{likertq}

\scale{Difficult to Find}

\begin{likertq}{I encountered}
\likertitem{a hidden gem (\textbf{something} interesting but that I did not know yet, or overlooked until now)}
\likertitem{\textbf{something} that might have been more difficult to discover without the help of this platform}
\end{likertq}

\begin{likertq}{I encountered}
\likertitem{\textbf{only} content that I would have easily discovered without the platform}
\end{likertq}

\scale{Unexpected}

\begin{likertq}{I encountered}
\likertitem{\textbf{something} unexpected}
\likertitem{\textbf{something} surprising}
\end{likertq}

\begin{likertq}{I encountered}
\likertitem{\textbf{only} content that did not surprise me}
\likertitem{\textbf{only} content that I expected to find}
\end{likertq}

\scale{Uncovering}

\begin{likertq}{I encountered}
\likertitem{\textbf{something} that helped me realize something new}
\likertitem{\textbf{something} that I did not know could be useful to me}
\likertitem{\textbf{something} that made me aware about something I did not know about before}
\likertitem{\textbf{something} that made me notice details I did not notice before}
\end{likertq}

\subsection*{Refreshing vs.\ Boring}

\scale{Novelty}

\begin{likertq}{I encountered}
\likertitem{\textbf{at least one item} that was new to me}
\likertitem{\textbf{at least one item} I had not seen before}
\likertitem{\textbf{at least one item} I did not know about yet}
\end{likertq}

\begin{likertq}{I encountered}
\likertitem{\textbf{only} items I had seen before}
\likertitem{\textbf{only} items I knew about already}
\likertitem{\textbf{nothing} new to me}
\end{likertq}

\scale{Unusual}

\begin{likertq}{I encountered}
\likertitem{\textbf{something} different from the type of content I normally consume}
\likertitem{\textbf{something} outside my typical preferences}
\likertitem{\textbf{something} different from my usual preferences}
\end{likertq}

\begin{likertq}{I encountered}
\likertitem{\textbf{only} the same genre of content I normally consume}
\end{likertq}

\scale{Taste-Reincarnation}

\begin{likertq}{I encountered}
\likertitem{\textbf{something} nostalgic}
\likertitem{\textbf{something} that reconnected me to my past}
\end{likertq}

\begin{likertq}{I encountered}
\likertitem{\textbf{something} from my personal past that I prefer to avoid}
\likertitem{\textbf{something} that reduced my interest in something I once liked}
\end{likertq}

\subsection*{Enriching vs.\ Brainrot}

\scale{Intrigue}

\begin{likertq}{I encountered}
\likertitem{\textbf{something} that sparked my interest}
\likertitem{\textbf{something} that stimulated my curiosity}
\end{likertq}

\begin{likertq}{I encountered}
\likertitem{\textbf{only} content that left me indifferent}
\likertitem{\textbf{only} content that failed to grab my attention}
\end{likertq}

\scale{Energy}

\begin{likertq}{I encountered}
\likertitem{\textbf{something} that recharged me}
\likertitem{\textbf{something} that gave me energy}
\end{likertq}

\begin{likertq}{I encountered}
\likertitem{\textbf{something} that lowered my energy}
\likertitem{\textbf{something} that left me feeling like a couch potato (lazy and without energy).}
\end{likertq}

\scale{Inspiration}

\begin{likertq}{I encountered}
\likertitem{\textbf{something} that stimulated my creativity}
\likertitem{\textbf{something} that stimulated my imagination}
\likertitem{\textbf{something} that sparked new ideas}
\likertitem{\textbf{something} that inspired me}
\end{likertq}

\scale{Impact}

\begin{likertq}{I encountered}
\likertitem{\textbf{something} that might improve my life}
\likertitem{\textbf{something} that might have a positive impact on my life}
\end{likertq}

\begin{likertq}{I encountered}
\likertitem{\textbf{something} that might have a negative impact on my life}
\end{likertq}

\begin{likertq}{I encountered}
\likertitem{\textbf{only} content that made me feel like I was wasting my time}
\end{likertq}

\scale{Empowering}

\begin{likertq}{I encountered}
\likertitem{\textbf{something} positive for my self-esteem}
\end{likertq}

\begin{likertq}{I encountered}
\likertitem{\textbf{something} negative for my self-esteem}
\end{likertq}

\scale{Useful}

\begin{likertq}{I encountered}
\likertitem{\textbf{something} useful for a problem I was facing}
\likertitem{\textbf{something} helpful}
\likertitem{\textbf{something} that might prove useful someday}
\end{likertq}

\begin{likertq}{I encountered}
\likertitem{\textbf{something} that might be misleading}
\likertitem{\textbf{something} that might be harmful}
\end{likertq}

\scale{Resonance}

\begin{likertq}{I encountered}
\likertitem{\textbf{something} that moved me}
\likertitem{\textbf{something} meaningful to me}
\likertitem{\textbf{something} touching}
\end{likertq}

\begin{likertq}{I encountered}
\likertitem{\textbf{something} annoying}
\likertitem{\textbf{something} toxic}
\likertitem{\textbf{something} disturbing}
\end{likertq}

\begin{likertq}{I encountered}
\likertitem{\textbf{only} content that felt empty}
\likertitem{\textbf{only} content that felt superficial}
\end{likertq}

\end{document}